\documentclass[letter,11pt,  final, onecolumn]{IEEEtran}
\usepackage{fancyhdr}
\usepackage{amsmath,epsfig}
\usepackage{threeparttable}
\usepackage{epsf,epsfig}
\usepackage{amsmath}
\usepackage{amssymb}
\usepackage{amsfonts}
\usepackage{algorithmic}
\usepackage{algorithm}
\usepackage[noadjust]{cite}
\usepackage{dsfont}
\usepackage{subfigure}
\newcommand{\aftersect}{\vspace{0pt}}
\newcommand{\beforesect}{\vspace{0pt}}

\begin{document}
\newtheorem{theorem}{\it Theorem}
\newtheorem{acknowledgement}[theorem]{Acknowledgement}
\newtheorem{axiom}[theorem]{Axiom}
\newtheorem{case}[theorem]{Case}
\newtheorem{claim}[theorem]{Claim}
\newtheorem{conclusion}[theorem]{Conclusion}
\newtheorem{condition}[theorem]{Condition}
\newtheorem{conjecture}[theorem]{Conjecture}
\newtheorem{criterion}[theorem]{Criterion}
\newtheorem{definition}[theorem]{Definition}
\newtheorem{example}[theorem]{Example}
\newtheorem{exercise}[theorem]{Exercise}
\newtheorem{lemma}{Lemma}
\newtheorem{corollary}{Corollary}
\newtheorem{notation}[theorem]{Notation}
\newtheorem{problem}[theorem]{Problem}
\newtheorem{proposition}{Proposition}
\newtheorem{solution}[theorem]{Solution}
\newtheorem{summary}[theorem]{Summary}
\newtheorem{assumption}{Assumption}
\newtheorem{examp}{\bf Example}
\newtheorem{probform}{\bf Problem}
\def\remark{{\noindent \bf Remark:\hspace{0.5em}}}

\def\qed{$\Box$}
\def\QED{\mbox{\phantom{m}}\nolinebreak\hfill$\,\Box$}
\def\proof{\noindent{\emph{Proof:} }}
\def\poof{\noindent{\emph{Sketch of Proof:} }}
\def
\endproof{\hspace*{\fill}~\qed
\par
\endtrivlist\unskip}
\def\endproof{\hspace*{\fill}~\qed\par\endtrivlist\vskip3pt}

\def\E{\mathbb{E}}
\def\eps{\varepsilon}
\def\phi{\varphi}
\def\Lsp{{\boldsymbol L}}
\def\Bsp{{\boldsymbol B}}
\def\lsp{{\boldsymbol\ell}}
\def\Ltsp{{\Lsp^2}}
\def\Lpsp{{\Lsp^p}}
\def\Linsp{{\Lsp^{\infty}}}
\def\LtR{{\Lsp^2(\Rst)}}
\def\ltZ{{\lsp^2(\Zst)}}
\def\ltsp{{\lsp^2}}
\def\ltZt{{\lsp^2(\Zst^{2})}}
\def\ninN{{n{\in}\Nst}}
\def\oh{{\frac{1}{2}}}
\def\grass{{\cal G}}
\def\ord{{\cal O}}
\def\dist{{d_G}}
\def\conj#1{{\overline#1}}
\def\ntoinf{{n \rightarrow \infty }}
\def\toinf{{\rightarrow \infty }}
\def\tozero{{\rightarrow 0 }}
\def\trace{{\operatorname{trace}}}
\def\ord{{\cal O}}
\def\UU{{\cal U}}
\def\rank{{\operatorname{rank}}}
\def\acos{{\operatorname{acos}}}

\def\SINR{\mathrm{SINR}}
\def\SNR{\mathrm{SNR}}
\def\SIR{\mathrm{SIR}}

\setcounter{page}{1}

% Definitions
\newcommand{\eref}[1]{(\ref{#1})}
\newcommand{\fig}[1]{Fig.\ \ref{#1}}

% Bold lowercase
\def\bydef{:=}
\def\ba{{\mathbf{a}}}
\def\bb{{\mathbf{b}}}
\def\bc{{\mathbf{c}}}
\def\bd{{\mathbf{d}}}
\def\bee{{\mathbf{e}}}
\def\bff{{\mathbf{f}}}
\def\bg{{\mathbf{g}}}
\def\bh{{\mathbf{h}}}
\def\bi{{\mathbf{i}}}
\def\bj{{\mathbf{j}}}
\def\bk{{\mathbf{k}}}
\def\bl{{\mathbf{l}}}
\def\bm{{\mathbf{m}}}
\def\bn{{\mathbf{n}}}
\def\bo{{\mathbf{o}}}
\def\bp{{\mathbf{p}}}
\def\bq{{\mathbf{q}}}
\def\br{{\mathbf{r}}}
\def\bs{{\mathbf{s}}}
\def\bt{{\mathbf{t}}}
\def\bu{{\mathbf{u}}}
\def\bv{{\mathbf{v}}}
\def\bw{{\mathbf{w}}}
\def\bx{{\mathbf{x}}}
\def\by{{\mathbf{y}}}
\def\bz{{\mathbf{z}}}
\def\b0{{\mathbf{0}}}

% Bold capital letters
\def\bA{{\mathbf{A}}}
\def\bB{{\mathbf{B}}}
\def\bC{{\mathbf{C}}}
\def\bD{{\mathbf{D}}}
\def\bE{{\mathbf{E}}}
\def\bF{{\mathbf{F}}}
\def\bG{{\mathbf{G}}}
\def\bH{{\mathbf{H}}}
\def\bI{{\mathbf{I}}}
\def\bJ{{\mathbf{J}}}
\def\bK{{\mathbf{K}}}
\def\bL{{\mathbf{L}}}
\def\bM{{\mathbf{M}}}
\def\bN{{\mathbf{N}}}
\def\bO{{\mathbf{O}}}
\def\bP{{\mathbf{P}}}
\def\bQ{{\mathbf{Q}}}
\def\bR{{\mathbf{R}}}
\def\bS{{\mathbf{S}}}
\def\bT{{\mathbf{T}}}
\def\bU{{\mathbf{U}}}
\def\bV{{\mathbf{V}}}
\def\bW{{\mathbf{W}}}
\def\bX{{\mathbf{X}}}
\def\bY{{\mathbf{Y}}}
\def\bZ{{\mathbf{Z}}}

% mathbb Bold capital letters
\def\mA{{\mathbb{A}}}
\def\mB{{\mathbb{B}}}
\def\mC{{\mathbb{C}}}
\def\mD{{\mathbb{D}}}
\def\mE{{\mathbb{E}}}
\def\mF{{\mathbb{F}}}
\def\mG{{\mathbb{G}}}
\def\mH{{\mathbb{H}}}
\def\mI{{\mathbb{I}}}
\def\mJ{{\mathbb{J}}}
\def\mK{{\mathbb{K}}}
\def\mL{{\mathbb{L}}}
\def\mM{{\mathbb{M}}}
\def\mN{{\mathbb{N}}}
\def\mO{{\mathbb{O}}}
\def\mP{{\mathbb{P}}}
\def\mQ{{\mathbb{Q}}}
\def\mR{{\mathbb{R}}}
\def\mS{{\mathbb{S}}}
\def\mT{{\mathbb{T}}}
\def\mU{{\mathbb{U}}}
\def\mV{{\mathbb{V}}}
\def\mW{{\mathbb{W}}}
\def\mX{{\mathbb{X}}}
\def\mY{{\mathbb{Y}}}
\def\mZ{{\mathbb{Z}}}

% Caligraphic capital letters
\def\cA{\mathcal{A}}
\def\cB{\mathcal{B}}
\def\cC{\mathcal{C}}
\def\cD{\mathcal{D}}
\def\cE{\mathcal{E}}
\def\cF{\mathcal{F}}
\def\cG{\mathcal{G}}
\def\cH{\mathcal{H}}
\def\cI{\mathcal{I}}
\def\cJ{\mathcal{J}}
\def\cK{\mathcal{K}}
\def\cL{\mathcal{L}}
\def\cM{\mathcal{M}}
\def\cN{\mathcal{N}}
\def\cO{\mathcal{O}}
\def\cP{\mathcal{P}}
\def\cQ{\mathcal{Q}}
\def\cR{\mathcal{R}}
\def\cS{\mathcal{S}}
\def\cT{\mathcal{T}}
\def\cU{\mathcal{U}}
\def\cV{\mathcal{V}}
\def\cW{\mathcal{W}}
\def\cX{\mathcal{X}}
\def\cY{\mathcal{Y}}
\def\cZ{\mathcal{Z}}
\def\cd{\mathcal{d}}
\def\Mt{M_{t}}
\def\Mr{M_{r}}
%% my defs
\def\O{\Omega_{M_{t}}}
\newcommand{\figref}[1]{{Fig.}~\ref{#1}}
\newcommand{\tabref}[1]{{Table}~\ref{#1}}

%% From Kaibin
\newcommand{\var}{\mathrm{Var}}
\newcommand{\fb}{\tx{fb}}
\newcommand{\nf}{\tx{nf}}
\newcommand{\BC}{\tx{(bc)}}
\newcommand{\MAC}{\tx{(mac)}}
\newcommand{\Pout}{P_{\tx{out}}}
\newcommand{\nnn}{\nn\\}
\newcommand{\FB}{\tx{FB}}
\newcommand{\TX}{\tx{TX}}
\newcommand{\RX}{\tx{RX}}
\renewcommand{\mod}{\tx{mod}}
\newcommand{\m}[1]{\mathbf{#1}}
\newcommand{\td}[1]{\tilde{#1}}
\newcommand{\sbf}[1]{\scriptsize{\textbf{#1}}}
\newcommand{\stxt}[1]{\scriptsize{\textrm{#1}}}
\newcommand{\suml}[2]{\sum\limits_{#1}^{#2}}
\newcommand{\sumlk}{\sum\limits_{k=0}^{K-1}}
\newcommand{\eqhsp}{\hspace{10 pt}}
\newcommand{\tx}[1]{\texttt{#1}}
\newcommand{\Hz}{\ \tx{Hz}}
\newcommand{\sinc}{\tx{sinc}}
\newcommand{\tr}{\mathrm{tr}}
\newcommand{\diag}{\mathrm{diag}}
\newcommand{\MAI}{\tx{MAI}}
\newcommand{\ISI}{\tx{ISI}}
\newcommand{\IBI}{\tx{IBI}}
\newcommand{\CN}{\tx{CN}}
\newcommand{\CP}{\tx{CP}}
\newcommand{\ZP}{\tx{ZP}}
\newcommand{\ZF}{\tx{ZF}}
\newcommand{\SP}{\tx{SP}}
\newcommand{\MMSE}{\tx{MMSE}}
\newcommand{\MINF}{\tx{MINF}}
\newcommand{\RC}{\tx{MP}}
\newcommand{\MBER}{\tx{MBER}}
\newcommand{\MSNR}{\tx{MSNR}}
\newcommand{\MCAP}{\tx{MCAP}}
\newcommand{\vol}{\tx{vol}}
\newcommand{\ah}{\hat{g}}
\newcommand{\tg}{\tilde{g}}
\newcommand{\teta}{\tilde{\eta}}
\newcommand{\heta}{\hat{\eta}}
\newcommand{\uh}{\m{\hat{s}}}
\newcommand{\eh}{\m{\hat{\eta}}}
\newcommand{\hv}{\m{h}}
\newcommand{\hh}{\m{\hat{h}}}
\newcommand{\Po}{P_{\mathrm{out}}}
\newcommand{\Poh}{\hat{P}_{\mathrm{out}}}
\newcommand{\Ph}{\hat{\gamma}}
\newcommand{\mat}[1]{\begin{matrix}#1\end{matrix}}
\newcommand{\ud}{^{\dagger}}
\newcommand{\C}{\mathcal{C}}
\newcommand{\nn}{\nonumber}
\newcommand{\nInf}{U\rightarrow \infty}

\title{ \setlength{\baselineskip}{30pt} Performance of Orthogonal Beamforming for SDMA with Limited Feedback}
\author{\large \setlength{\baselineskip}{20pt}Kaibin Huang, Jeffrey G.
Andrews, and Robert W. Heath, Jr
\thanks{ The authors are with Wireless Networking and Communications Group, Department of Electrical and Computer Engineering, The University of Texas at Austin, 1 University Station C0803, Austin, TX 78712. Email: huangkb@mail.utexas.edu, \{jandrews, rheath\}@ece.utexas.edu. Kaibin Huang is the recipient of the University Continuing Fellowship from The University of Texas at Austin. This work is funded by the DARPA IT-MANET program under the grant W911NF-07-1-0028, and the National Science Foundation under grants CCF-514194 and CNS-435307. The results in this paper were presented
in part at the IEEE Int. Conf. Acoust., Speech and Sig. Proc., Apr.
2007.}} \markboth{Accepted for publication in IEEE Transactions on Vehicular Technology, Revised on \today}{}
\maketitle

%\begin{keywords}\aftersect
%Array Signal Processing, Space Division Multiplexing, Multiuser
%Channels, Diversity Methods, Feedback Communication, Broadcast
%Channels, Scheduling
%\end{keywords}

\begin{abstract}\setlength{\baselineskip}{15pt}
On the multi-antenna broadcast channel, the spatial degrees of freedom support simultaneous transmission to multiple users. The optimal multiuser transmission, known as dirty paper coding, is not directly realizable. Moreover, close-to-optimal solutions such as Tomlinson-Harashima precoding are sensitive to CSI inaccuracy. This paper considers a more practical design called per user unitary and rate control (PU2RC), which has been proposed for emerging cellular standards. PU2RC supports multiuser simultaneous transmission, enables limited feedback, and is capable of exploiting multiuser diversity. Its key feature is an orthogonal beamforming (or precoding) constraint, where each user selects a beamformer (or precoder) from a codebook of multiple orthonormal bases. In this paper, the asymptotic throughput scaling laws for PU2RC with a large user pool are derived for different regimes of the signal-to-noise ratio (SNR). In the multiuser-interference-limited regime, the throughput of PU2RC is shown to scale logarithmically with the number of users. In the normal SNR and noise-limited regimes, the throughput is found to scale double logarithmically with the number of users and also linearly with the number of antennas at the base station. In addition, numerical results show that PU2RC achieves higher throughput and is more robust against CSI quantization errors than the popular alternative
of zero-forcing beamforming if the number of users is sufficiently large.
\end{abstract}

\beforesect\beforesect\section{Introduction}\aftersect\label{Section:Introd}
In multi-antenna broadcast channels, simultaneous transmission to
multiple users, known as multiuser multiple-input-multiple-output (MIMO) or \emph{space division multiple access}
(SDMA), is capable of achieving much higher throughput than other
multiple-access schemes such as \emph{time division multiple access}
(TDMA) \cite{Viswanath:SumCapBroadcastChan:2003}. Due to this
advantage, SDMA has been  recently included in the IEEE 802.16e
standard \cite{IEEE802-16e}, and has been proposed for the emerging 3GPP long
term evolution (LTE) standard
\cite{3GPP-LTE,LTEMIMO:06,PU2RC:06,LTEZFMIMO:06}. While the optimal SDMA
strategy is known, \emph{dirty paper coding}
\cite{Costa:WriteDirtyPaper:83} is non-causal and hence not
directly realizable. Moreover, close-to-optimal techniques such as Tomlinson-Harashima precoding and vector perturbation are sensitive to CSI inaccuracy \cite{Yu:TrelliConvPrecodInterSub:2005, SpencerETAL:introsdma:04}. More  practical SDMA algorithms are based on
transmit beamforming, including
zero forcing
\cite{Choi:TXPreprocMuMIMO:2004,Wong:ChanDiagMuMIMO:2003,SpencerSwindleETAL:ZFsdma:2004,Dimic:LowCompDLBeamfMaxCap:2004},
a signal-to-interference-plus-noise-ratio (SINR) constraint
\cite{SchubertBoche:solsdma:04},
 minimum mean squared error (MMSE) \cite{Serbetli:TxRxOptimMuMIMO:2004},
and channel decomposition \cite{Choi:ComplementBeamf:2006}. These
SDMA algorithms can be combined with multiuser scheduling to
further increase the throughput by exploiting \emph{multiuser
diversity}, which refers to scheduling only a subset of users with
good channels for each transmission
\cite{Knopp:InfoCapPwrCtrlMu:1995,SwannackWorenell:BroadcastChanLimitedCSI:2005,
YooGoldsmith:OptimBroadcastZeroForcingBeam:2006,
YooJindal:FiniteRateBroadcastMUDiv:2007,ChoiForenza:OppSDMABeamSel:06,SharifHassibi:CapMIMOBroadcastPartSideInfo:Feb:05,
ShenChen:LowCompUsrSelBD:2006}. Both scheduling and beamforming in a
SDMA system require channel state information (CSI) at the base
station. Unfortunately, CSI feedback from each user potentially
incurs excessive overhead because of the multiplicity of channel
coefficients. Therefore, this paper focuses on SDMA that supports
efficient CSI feedback and uses CSI for joint beamforming and
scheduling.

\beforesect\subsection{Related Work and
Motivation}\aftersect\label{RelateWork} In this paper, we consider a
practical scenario where partial CSI is acquired by the base station
through quantized CSI feedback, known as \emph{limited feedback}
\cite{LovHeaETAL:WhatValuLimiFee:Oct:2004}. Quantized CSI feedback
for point-to-point communications has been extensively studied
recently (see e.g.
\cite{LovHeaETAL:WhatValuLimiFee:Oct:2004,LoveHeathBook} and the
references therein). The effects of CSI quantization on a SDMA
system have been investigated in
\cite{DingLove:SubspaceFbBroadcastChan:2005,Jindal:MIMOBroadcastFiniteRateFeedback:06,YooJindal:FiniteRateBroadcastMUDiv:2007}.
The key result of
\cite{YooJindal:FiniteRateBroadcastMUDiv:2007} is that the number of
CSI feedback bits can be reduced by exploiting multiuser diversity. In \cite{DingLove:SubspaceFbBroadcastChan:2005}, combined quantized
CSI feedback and zero-forcing dirty paper coding are shown to attain
most of the capacity achieved by perfect CSI feedback. In
\cite{Jindal:MIMOBroadcastFiniteRateFeedback:06}, it is shown that
for a small number of users the number of CSI feedback bits must
increase with the signal-to-noise ratio (SNR) to ensure that the
throughput  grows with SNR.

This paper addresses joint beamforming and scheduling for SDMA
systems to maximize throughput, assuming backlogged users. A similar scenario but with bursty data and the
objective of meeting quality-of-service (QoS) for different users is
addressed in \cite{YinLiu02:PerfSDMASchedule} and references
therein.  The optimal approach for our full-queue scenario involves
an exhaustive search, where for each possible subset of users a
corresponding set of beamforming vectors is designed using
algorithms such as that proposed in \cite{SchubertBoche:solsdma:04}. The main drawback of the optimal
approach is its complexity, which increases exponentially with the
number of users. This motivates the designs of more efficient SDMA
algorithms.

In \cite{SharifHassibi:CapMIMOBroadcastPartSideInfo:Feb:05}, a
practical SDMA algorithm, called \emph{opportunistic SDMA} (OSDMA), is
proposed, which supports low-rate beamforming feedback and satisfies
the orthogonal beamforming constraint. As shown in
\cite{SharifHassibi:CapMIMOBroadcastPartSideInfo:Feb:05}, for a
large number of users, an arbitrary set of orthogonal beamforming
vectors ensures that the throughput  increases with the number of
users at the optimal rate. Nevertheless, for a small number of
users, such arbitrary beamforming vectors are highly sub-optimal due
to excessive interference between scheduled users. To reduce
multiuser interference caused by sub-optimal beamforming vectors,
an extension of OSDMA, called \emph{OSDMA with beam selection}
(OSDMA-S), is proposed in \cite{ChoiForenza:OppSDMABeamSel:06},
 where each mobile iteratively selects beamforming vectors
broadcast by the base station and sends back its choices. Due to
distributed beam selection, numerous iterations of broadcast and
feedback are required for implementing OSDMA-S, which incurs
significant downlink overhead and feedback delay. As a result, the
throughput gains of OSDMA-S over OSDMA are marginal.

An alternative beamforming SDMA algorithm is proposed in
\cite{YooJindal:FiniteRateBroadcastMUDiv:2007}, referred to as
ZF-SDMA, where feedback CSI is quantized using the \emph{random vector
quantization} (RVQ) algorithm \cite{YeungLove:RandomVQBeamf:05,
Jindal:MIMOBroadcastFiniteRateFeedback:06} and greedy-search
scheduling is performed prior to zero-forcing beamforming. A design similar to ZF-SDMA \cite{LTEZFMIMO:06} has been proposed to
the emerging 3GPP-LTE standard \cite{3GPP-LTE}, which is the latest cellular communication standard. The
drawback of ZF-SDMA is its lack of robustness against CSI
inaccuracy due to the separate designs of the limited feedback,
scheduling and beamforming sub-algorithms.

In industry, SDMA with orthogonal beamforming, under the name
\emph{per user unitary and rate control} (PU2RC)
\cite{PU2RC:06}, has been proposed to the 3GPP-LTE standard. The main feature of PU2RC is limited feedback, where
multiuser precoders or beamformers are selected from a codebook of
multiple orthonormal bases. Based on limited feedback, PU2RC
supports SDMA, scheduling, and adaptive modulation and coding.
Because of its versatility and advanced features, PU2RC is one of the most promising solutions for high-speed downlink in 3GPP-LTE. The importance of
PU2RC for the next-generation wireless communication motivates the
investigation of its performance in this paper.

In this paper, we consider a simplified PU2RC system where scheduled
users have single data streams, which are separated by orthogonal
beamformers. In this case, PU2RC generalizes OSDMA
\cite{SharifHassibi:CapMIMOBroadcastPartSideInfo:Feb:05} by allowing
the beamforming codebook to contain more than one orthonormal basis.
Such a generalization complicates the performance analysis of PU2RC
because the resultant scheduler is more complicated. To be specific,
the scheduler has to select an orthonormal basis from the codebook
besides choosing a particular user for each codebook vector. Such a
challenge motivates our use of a new analytical tool, namely
\emph{uniform convergence in the weak law of large
numbers} \cite{VapCher:UniformConvLawLargeNumb:2005},
for analyzing the throughput of PU2RC instead of \emph{extreme value
theory} as applied in
\cite{SharifHassibi:CapMIMOBroadcastPartSideInfo:Feb:05}.

Theory of uniform convergence in the weak law of large
numbers is also applied in our previous work \cite{Huang:UplinkSDMALimFb:07} for analyzing the throughput of uplink SDMA with limited feedback. Despite using the same tool, the analysis in this paper differs from  \cite{Huang:UplinkSDMALimFb:07} due to differences between the uplink and downlink. Specifically, the received data signal for the downlink propagates through a single-user channel, but that for the uplink passes through multiuser channels. As a result, SINR feedback for downlink SDMA is infeasible for uplink SDMA, where SINR depends on multiuser CSI and is hence uncomputable at users. Consequently, downlink and uplink SDMA require different designs of scheduling algorithm. Thus, the joint beamforming and scheduling algorithm presented in this paper is not applicable for uplink SDMA. Interestingly, despite the differences between the uplink and downlink, the asymptotic throughput scaling laws for downlink SDMA as derived in this paper are found to be identical to those for uplink SDMA \cite{Huang:UplinkSDMALimFb:07}.

\beforesect\subsection{Contributions and
Organization}\aftersect\label{Contrib} The main contribution of this
paper is the analysis of the throughput scaling of PU2RC for an
asymptotically large number of users $U\rightarrow\infty$. Using the
theory of uniform convergence in the weak law of large numbers,
throughput scaling laws are derived for three regimes, namely the
\emph{normal SNR}, \emph{interference-limited} and the
\emph{noise-limited} regimes. In the normal SNR regime, both the
variance of noise and multiuser interference are comparable; in the
interference-limited  regime, multiuser interference dominates over noise; the
reverse exists in the noise-limited regime. Our main results are summarized as follows. In the
interference-limited regime, we show that the throughput scales
\emph{logarithmically} with $U$ but does not increase with the
number of transmit antennas $N_t$ at the base station. In both the normal SNR and noise-limited regimes, we show that the throughput scales \emph{double logarithmically} with $U$ and \emph{linearly} with $N_t$. This throughput scaling law shows that PU2RC achieves the optimal multiuser diversity gain as OSDMA in the normal SNR regime\footnote{The interference-limited and noise-limited regimes
have not been considered for OSDMA in
\cite{SharifHassibi:CapMIMOBroadcastPartSideInfo:Feb:05}}. Thereby,
this result contradicts the intuition that using multiple
orthonormal bases in the codebook splits the user pool and hence
reduces the multiuser diversity gain. Using Monte
Carlo simulations, the asymptotic throughput scaling laws are
also found to hold in the non-asymptotic regime where $U$ is
finite.

The asymptotic throughput analysis for PU2RC provides several
guidelines for designing the scheduler to ensure optimal throughput
scaling. First, in the interference-limited regime, scheduling
should use the criterion of minimum quantization error. Second, in
the normal SNR regime, scheduled users should have both large channel power and small quantization errors. Third, in the noise-limited
feedback, scheduling should select users with large channel power
while the quantization error is a less important scheduling
criterion.

Numerical results are presented for evaluating the throughput of
PU2RC  and also comparing PU2RC with ZF-SDMA. Several observations
are made. First, increasing the amount of CSI feedback (or the
codebook size) can decrease the throughput for PU2RC if the number
of users is small. Otherwise, more CSI feedback provides a
throughput gain. Second, PU2RC achieves higher throughput than
ZF-SDMA for large numbers of users but the reverse holds for
relatively small numbers of users. Third, decreasing the codebook
size causes a larger throughput loss for ZF-SDMA than that for
PU2RC.

The remainder of this paper is organized as follows. The system
model is described in Section~\ref{Section:Sys}. The sub-algorithms
of PU2RC for CSI  quantization, and joint beamforming and scheduling are
presented in Section~\ref{Section:Algo}. The asymptotic throughput
scaling of PU2RC is analyzed in Section~\ref{Section:Anlysis}. The
performance of PU2RC is evaluated using Monte Carlo simulation in
Section~\ref{Section:Numerical}, followed by concluding remarks in
Section~\ref{Section:Conclusion}.

\beforesect\section{System Model}
\begin{figure}
\centering
  \includegraphics[width=13cm]{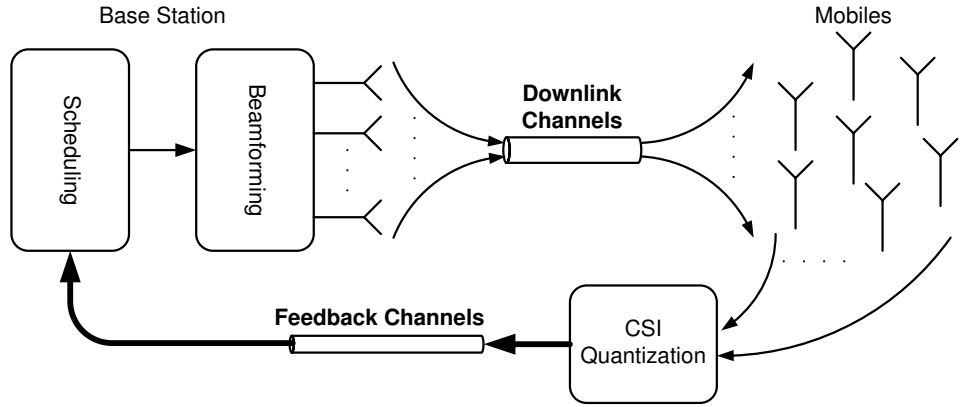}\\
\caption{Downlink system with limited feedback }\label{Fig:DLSYS}
\end{figure}

\aftersect\label{Section:Sys} The
downlink or broadcast system illustrated in Fig.~\ref{Fig:DLSYS} is
described as follows. The base station with $N_t$ antennas transmits
data simultaneously to $N_t$ active users chosen from a total of $U$
users, each with one receive antenna. The base station separates the
multiuser data streams by beamforming, i.e. assigning a beamforming
vector to each of the $N_t$ active users. The beamforming vectors
$\{\bw_n\}_{n=1}^{N_t}$ are selected from multiple sets of unitary
orthogonal vectors following the beam and user selection algorithm
described in Section~\ref{Section:BeamSel}. Equal power allocation
over scheduled users is considered\footnote{Note that equal power
allocation is close to the optimal water-filling method if
scheduled users all have high SINR.}. The received signal of the $u$th
scheduled user is expressed as
\begin{equation}\label{Eq:Sys}
    y_u = \sqrt{\frac{P}{N_t}}\bh^\dagger_u\suml{n\in\mathcal{A}}{}\bw_nx_n + \nu_u, \quad u \in \mathcal{A},
\end{equation}
where we use the following notation

\begin{minipage}[t]{8.25cm}
  \begin{description}
  \item[$N_t$] number of transmit antennas and also number of
  scheduled users;
  \item[$\bh_u$]  ($N_t\times 1$ vector) downlink channel;
    \item[$x_u$] transmitted symbol with $E[|x_u|^2] = 1$;
  \item[$y_u$] received symbol;
  \item[$\dagger$] conjugate transpose matrix operation;
\end{description}

\end{minipage}
\hspace{3mm}
\begin{minipage}[t]{8.25cm} %\begin{figure}
\begin{description}
  \item[$\bw_u$] ($N_t\times 1$ vector) beamforming vector with
  $\|\bw_u\|^2=1$;
  \item[$\mathcal{A}$] The index set of scheduled users;
  \item[$P$] transmission power; and
  \item[$\nu_u$] AWGN sample with $\nu_u\sim\mathcal{CN}(0,1)$.
\end{description}
\end{minipage}
\vspace{5pt}

For the purpose of asymptotic analysis of PU2RC, we make the
following assumption:
\begin{assumption}\label{AS:DLChan}\emph{
The downlink channel $\bh_u$ $\forall \ u=1, 2, \cdots, U$ is an
i.i.d. vector with $\mathcal{CN}(0,1)$ coefficients.}
\end{assumption}
Given this assumption commonly made in the literature of SDMA and
multiuser diversity
\cite{SharifHassibi:CapMIMOBroadcastPartSideInfo:Feb:05,
ChoiForenza:OppSDMABeamSel:06,
Jindal:MIMOBroadcastFiniteRateFeedback:06,
SwannackWorenell:BroadcastChanLimitedCSI:2005,
YooGoldsmith:OptimBroadcastZeroForcingBeam:2006}, the channel
direction vector $\bh_u/\|\bh_u\|$ of each user follows a uniform
distribution. Assumption~\ref{AS:DLChan}
greatly simplifies the throughput analysis of PU2RC in
Section~\ref{Section:Anlysis} but has no effect on the PU2RC
algorithms in Section~\ref{Section:Algo}. Assumption~\ref{AS:DLChan}
is valid for the scenario where wireless channels have rich
scattering and users encounter equal path loss. Throughput
analysis for a more complicated channel model is a topic for future
investigation.

\beforesect\section{Algorithms}\aftersect\label{Section:Algo} In
this section, we propose the algorithms for PU2RC including (i)
limited feedback by the mobiles and (ii) joint beamforming and
scheduling at the base station. The principles for these algorithms
have been described in the proposal of PU2RC \cite{PU2RC:06} even
though their details are not provided therein. The algorithms
presented in the following sections are tailored for the system model in
Section~\ref{Section:Sys}. The following discussion on algorithms
serves two purposes: (i) to elaborate the operation of PU2RC and
(ii) to establish an analytical model for the  asymptotic throughput
analysis in Section~\ref{Section:Anlysis}.

\subsection{Limited
Feedback}\aftersect\label{Section:Quant} Without loss of generality,
the discussion in this section focuses  on the $u$th user and the
same algorithm for CSI quantization is used by other users. For
simplicity, we make the following assumption
\begin{assumption}\label{AS:RXCSI}
\emph{The $u$th user has perfect CSI $\bh_u$.}
\end{assumption}
This assumption allows us to neglect the channel estimation error at the
$u$th mobile. For convenience, the CSI, $\bh_u$, is decomposed into
two components: the \emph{gain} and the \emph{shape}.
Hence,\vspace{-10pt}
\begin{equation}\label{Eq:ChanDecomp}
    \bh_u = g_u\bs_u,\quad u = 1, \cdots, U,
\end{equation}
where $g_u = \|\bh_u\|$ is the gain and $\bs_u = \bh_u/\|\bh_u\|$ is
the shape. The $u$th user quantizes and sends back to the base
station two quantities: the \emph{channel shape} and the
SINR.

The channel shape $\bs_u$ is quantized using a codebook-based
quantizer \cite{GerGra:VectQuanSignComp:92} with a codebook
comprised of multiple sets of orthonormal vectors in
$\mathds{C}^{N_t}$. Let $\mathcal{F}$ denote the codebook,
$\mathcal{V}^{(m)}$ the $m$th orthonormal set in the codebook,  and
$M$ the number of such sets. Thus,
$\mathcal{F}=\bigcup_{m=1}^M\mathcal{V}^{(m)}$ and the codebook size is $N = |\mathcal{F}|=MN_t$. For our design, the
$M$ orthonormal bases of $\mathcal{F}$ are generated randomly and
independently using a method such as that in
\cite{Zyczkowski:RandUnitMatrices:94}. Following
\cite{LovHeaETAL:GrasBeamMultMult:Oct:03} and \cite{MukSabETAL:BeamFiniRateFeed:Oct:03},
the quantized channel shape, represented by  $\uh_u$, is the member
of $\mathcal{F}$ that forms the smallest angle with the channel
shape $\bs_u$. Mathematically,
\begin{equation}\label{Eq:Quant}
    \uh_u =
    \arg\min_{\bv\in\mathcal{F}}d(\bv, \bs_u),
\end{equation}
where the distortion function $d(\bv, \bs_u)$ is given as
\begin{equation}\label{Eq:Dist}
d(\bv, \bs_u) =     1- \left|\bv^\dagger\bs_u\right|^2= \sin^2(\angle(\bv, \bs_u)).
\end{equation}
It follows that the \emph{quantization error}
can be defined as $\epsilon = \sin^2(\angle(\uh_u, \bs_u))$. It is
clear that $\epsilon = 0$ if $|\uh_u^\dagger\bs_u|=1$ and  $\epsilon
= 1$ if $\uh_u\perp\bs_u$.

The quantized channel shape $\uh_u$ is sent back to the base station
through a finite-rate feedback channel
\cite{LovHeaETAL:WhatValuLimiFee:Oct:2004,LovHeaETAL:GrasBeamMultMult:Oct:03}.
Since the quantization codebook $\mathcal{F}$ can be known \emph{a
priori} to both the base station and mobiles, only the index of
$\uh_u$ needs to be sent back. Therefore, the number of feedback
bits per user for quantized channel shape feedback is $\log_2N$ since
$|\mathcal{F}|=N$. The number of additional bits required for SINR
feedback is discussed in Section~\ref{Section:Numerical}.

Besides the channel shape, the $u$th user also sends back to the
base station the SINR, which serves as a channel quality indicator.
For orthogonal beamforming, the SINR is given as
\cite{YooJindal:FiniteRateBroadcastMUDiv:2007}
\begin{equation}\label{Eq:SINR:Exp}
    \SINR_u =
    \frac{\gamma\rho_u(1-\epsilon_u)}{1+\gamma\rho_u\epsilon_u},
\end{equation}
where $\gamma=\frac{P}{N_t}$ is the SNR, $\epsilon_u$ the CSI quantization error, and $\rho_u=\|\bh_u\|^2$  the channel power. Since the SINR is a scalar and requires much fewer
feedback bits than the channel shape, we make the following
assumption:
\begin{assumption}\label{AS:ChanGain}\emph{
The $\SINR_u$ is perfectly known to the base station through
feedback.}
\end{assumption}
The same assumption is also made in
\cite{SharifHassibi:CapMIMOBroadcastPartSideInfo:Feb:05,ChoiForenza:OppSDMABeamSel:06}.
The effect of SINR quantization on the throughput  is shown to be
insignificant using numerical results in
Section~\ref{Section:Numerical}.

\beforesect\beforesect\subsection{Joint Scheduling and
Beamforming}\aftersect\label{Section:BeamSel} This section focuses on the joint scheduling and beamforming algorithm designed based on the principles of PU2RC. Having collected
quantized CSI from all $U$ users\footnote{For simplicity, we assume
that the number of feedback bits per user is limited but not the
total number of feedback bits from all users. Nevertheless, the sum
feedback from all users can be reduced by allowing only a small
subset of users for feedback, which is an topic addressed in a
separate paper \cite{Huang:SDMASumFbRate:06}.}, the base station
schedules $N_t$ users for transmission and computes their
beamforming vectors. To maximize the throughput, $N_t$ scheduled
users must be selected through an exhaustive search, which is
infeasible for a large user pool. Therefore, we adopt a simpler
joint scheduling and beamforming algorithm. In brief, this algorithm
schedules a subset of users with orthogonal quantized channel shapes,
and furthermore applies these channel shapes as the scheduled users'
beamforming vectors.

The joint scheduling and beamforming algorithm is elaborated as
follows. First, each member of the codebook $\mathcal{F}$, which is
a potential beamforming vector, is assigned a user with the maximum
SINR. Consider an arbitrary vector, for instance $\bv^{(m)}_n$,
which is the $n$th member of the $m$th orthonormal subset
$\mathcal{V}^{(m)}$ of the codebook $\mathcal{F}$. This vector can
be the quantized channel shapes of multiple users, whose indices are
grouped in a set defined as $\mathcal{I}^{(m)}_n = \left\{1\leq
u\leq U: \uh_u=\bv^{(m)}_n\right\}$ where $\uh_u$ is the $u$th
user's quantized channel shape given in \eqref{Eq:Quant}. From
\eqref{Eq:Quant}, $\mathcal{I}^{(m)}_n$ can be equivalently defined
as
\begin{equation}\label{Eq:I-set}
\mathcal{I}^{(m)}_n = \left\{1\leq u\leq U\mid d\left(\bs_u, \bv^{(m)}_n\right) < d\left(\bs_u, \bv\right)\
\forall \ \bv\in\mathcal{F}\  \text{and} \
\bv\neq\bv^{(m)}_n\right\}.
\end{equation}
Among the users in $\mathcal{I}^{(m)}_n$,
$\bv^{(m)}_n$ is associated with the one providing the maximum SINR,
which is feasible since the SNRs are known to the base station
through feedback. The index $\left(i^{(m)}_n\right)$ and SINR $\left(\xi^{(m)}_n\right)$ of
this user associated with $\bv^{(m)}_n$ can be written as
\begin{equation}
    i^{(m)}_n = \arg\max_{u\in\mathcal{I}^{(m)}_n}\SINR_u \quad \text{and} \quad
    \xi^{(m)}_n = \max_{u\in\mathcal{I}^{(m)}_n}\SINR_u,\label{Eq:UserSelect}
\end{equation}
where the index set $\mathcal{I}^{(m)}_n$ and the function
$\SINR_u$ are expressed respectively in \eqref{Eq:I-set} and
    \eqref{Eq:SINR:Exp}.
In the event that $\mathcal{I}^{(m)}_n=\emptyset$, the vector
    $\bv^{(m)}_n$ is associated with no user and the maximum SINR
    $\xi^{(m)}_n$ in \eqref{Eq:UserSelect} is set as zero.
Second, the orthonormal subset of the codebook
that maximizes throughput  is
chosen, whose index is $m^\star=\arg\max_{1\leq m\leq M}\sum_{n=1}^{N_t}\log\left(1+    \xi^{(m)}_n\right)$. Thereby, the users associated with this chosen subset, specified by the indices $\left\{i^{(m^\star)}_n\mid 1\leq n\leq N_t\right\}$, are
scheduled for simultaneous transmission using beamforming vectors from the $(m^\star)$th orthonormal subset.

The above scheduling algorithm does not guarantee that the number
of scheduled users is equal to $N_t$, the spatial degrees of
freedom. For a small user pool, the number of scheduled users is
smaller than $N_t$. This is desirable because it is unlikely to find
$N_t$ simultaneous users with close-to-orthogonal channels in a
small user pool. In this case, having fewer scheduled users than
$N_t$ reduces interference and leads to higher throughput. As the
total number of users increases, the number of scheduled users
converges to $N_t$. Numerical results on the average number of
scheduled users for PU2RC are presented in
Section~\ref{Section:Numerical}.

 Based on the preceding algorithm for joint beamforming and scheduling,
the ergodic throughput for PU2RC is given as
\begin{equation}\label{Eq:Thput:Exp}
R = \E\left[\max_{1\leq m \leq
M}\sum_{n=1}^{N_t}\log\left(1+\max_{u\in\mathcal{I}^{(m)}_n}\SINR_u\right)\right]
\end{equation}
where $\SINR_u$ is given in \eqref{Eq:SINR:Exp}. The scaling of $R$
with the number of users $U$ as $U\rightarrow\infty$ is analyzed in
Section~\ref{Section:Anlysis}.

\section{Asymptotic Throughput
Scaling}\label{Section:Anlysis} In this section, we derive the
scaling laws of the PU2RC throughput for an asymptotically large
number of users. Auxiliary results required in
the analysis are first presented in Section~\ref{Section:Auxiliary}. Three
SNR regimes, namely \emph{normal}, \emph{interference-limited}, and \emph{noise-limited} regimes, are
considered in Section~\ref{Section:NorRegime} to \ref{Section:NoiseRegime}, respectively. Finally, numerical results showing how the asymptotic throughput scaling laws apply in the non-asymptotic regime are presented in Section~\ref{Section:NonAsympRegime}.
The asymptotic throughput scaling laws derived in this section for downlink SDMA are observed to be identical to those for uplink SDMA \cite{Huang:UplinkSDMALimFb:07}. This suggests duality between uplink and downlink SDMA in terms of asymptotic throughput.

\subsection{Auxiliary Results}\label{Section:Auxiliary}
Two auxiliary results are provided in this section. In
Section~\ref{Section:LawLargeNumb}, the theory of uniform
convergence in the weak law of large numbers is discussed, which is
an important tool for the subsequent asymptotic throughput analysis.
The other useful result related to the channel-shape quantization
error is presented in Section~\ref{Section:QuantErr}.

\subsubsection{Uniform Convergence in the weak law of large numbers}\label{Section:LawLargeNumb} In this section, a lemma on the
uniform convergence in the weak law of large numbers
\cite{VapCher:UniformConvLawLargeNumb:2005} is obtained by
generalizing \cite[Lemma~4.8]{GuptaKumar:CapWlssNetwk:2000} from
$\mathds{R}^3$ to $\mathds{C}^{N_t}$. This lemma is useful for
analyzing the number of users whose channel shapes lie in one of a
set of congruent disks on the surface of a unit hyper-sphere in
$\mathds{C}^{N_t}$.

%------------------- Lemma: Gupta and Kumar ----------------------
\begin{lemma}[Gupta and Kumar]\label{Lem:LargeNum} \emph{Consider $U$ random points uniformly distributed on
the surface of a unit hyper-sphere in $\mathds{C}^{N_t}$  and $N$
disks on the sphere surface that have equal volume denoted as $A$.
Let $T_n$ denote the number of points belong to the $n$th disk. For
every $\tau_1,\tau_2> 0$
\begin{equation}
    \Pr\left(\sup_{1\leq n\leq N }\left|\frac{T_n}{U}-A\right|\leq
    \tau_1\right)>
1-\tau_2, \quad U \geq U_o
\end{equation}
where
\begin{equation}
U_o = \max\left\{\frac{3}{\tau_1}\log\frac{16c}{\tau_2},
\frac{4}{\tau_1}\log\frac{2}{\tau_2} \right\}\label{Eq:Uo}
\end{equation}
and  $c$ is a constant.}
\end{lemma}
\begin{proof}
See Appendix~\ref{App:LargeNum}.
\end{proof}

\subsubsection{Quantization Error of Channel Shape} \label{Section:QuantErr}
The complementary
CDF of the CSI quantization error $\epsilon $ is analyzed as
follows. As defined in Section~\ref{Section:Quant}, $\epsilon =
\sin^2(\angle(\uh, \bs))$ where $\bs$ and $\uh$ are the original and
the quantized channel shapes of an arbitrary user. From the
quantization function in \eqref{Eq:Quant}, the complementary CDF of
$\epsilon$ is
\begin{equation}\label{Eq:CDF:Def}
\Pr(\epsilon\geq \delta) = \Pr\left(\bs\notin
\bigcup_{\bv\in\mathcal{F}}B_{\delta}(\bv)\right),
\end{equation}
where $0\leq \delta\leq 1$ and $B_{\delta}(\bv) = \left\{\bs\in
\mathds{O}^{N_t}: |\bs^\dagger\bv|^2 \leq \delta\right\} $ is a
sphere cap on the unit sphere $\mathds{O}^{N_t}$. The CDF of
$\epsilon$ for $0\leq \delta\leq \tfrac{1}{2}$ has the simple
expression as given in the following lemma, but the derivation of
CDF for $\tfrac{1}{2}\leq \delta\leq 1$ is difficult because the
sphere caps $\{B_{\delta}(\bv): \bv \in\mathcal{F} \}$ overlap.
\begin{lemma}\label{Lem:EpsCDF}
\emph{The complementary CDF of $\epsilon$, $\Pr(\epsilon \geq
\delta)$,   for $0\leq \delta\leq \tfrac{1}{2}$ is given as
\begin{equation}
\Pr(\epsilon \geq \delta) = \left[1- N_t\delta^{N_t-1}\right]^M,
\quad 0\leq \delta \leq \tfrac{1}{2},\label{Eq:EpsCDF}
\end{equation}
where $M$ is the number of orthonormal bases in the quantization
codebook $\mathcal{F}$. In addition, $\Pr(\epsilon \geq \delta)\leq
\left(1- \delta^{N_t-1}\right)^M \ \forall \ 0\leq \delta \leq 1$. }
\end{lemma}
\begin{proof} See Appendix~\ref{App:EpsCDF}.\end{proof}

Next, the following lemma provides an upper-bound for the quantity
$\E[-\log \epsilon]$, which is useful for the throughput analysis in
the sequel. The derivation of this result uses
Lemma~\ref{Lem:EpsCDF} and
\cite[Lemma~4]{Jindal:MIMOBroadcastFiniteRateFeedback:06}.
\begin{lemma}\label{Lem:LogEps}\emph{Given a codebook of $M$
orthonormal bases, the following inequality holds
\begin{equation}
\frac{\log M }{(N_t-1)P_{\alpha}} + \frac{\log N_t}{N_t-1} \leq
\E\left[-\log\epsilon\right]\leq \frac{\log M +
1}{(N_t-1)P_{\alpha}} + \frac{\log N_t}{N_t-1}
\end{equation} where $\epsilon$ is the channel-shape quantization error and
\begin{equation}\label{Eq:Pa}
P_{\alpha}=    1-\left[1-N_t2^{-(N_t-1)}\right]^M.
\end{equation}}
\end{lemma}
\begin{proof} See Appendix~\ref{App:LogEps}.
\end{proof}

\subsection{Normal SNR Regime}\label{Section:NorRegime}
In this section, the throughput scaling law of PU2RC is analyzed for
the normal SNR regime, where the SINR and
throughput are given respectively in \eqref{Eq:SINR:Exp} and
\eqref{Eq:Thput:Exp}. As shown in the sequel, in the normal SNR regime,
the throughput of PU2RC scales double logarithmically with the
number of users and linearly with the number of antennas. This
throughput scaling law is identical to those for ZF-SDMA
\cite{YooJindal:FiniteRateBroadcastMUDiv:2007} and OSDMA
\cite{SharifHassibi:CapMIMOBroadcastPartSideInfo:Feb:05}. Therefore,
these algorithms all achieve optimal multiuser diversity gain.

The procedure for deriving the throughput scaling law for PU2RC is
to first obtain an upper-bound for the throughput scaling factor and
second prove its achievability. The achievability  proof uses
Lemma~\ref{Lem:LargeNum} on the uniform convergence in the weak law
of large numbers. The above procedure is also adopted for the
throughput analysis for other regimes in subsequent sections.

For the normal SNR regime,  the throughput scaling factor for PU2RC is
upper bounded as shown in the following lemma.
\begin{lemma}\label{Lem:ThputUB:NoSNR}\emph{
In the normal SNR regime, the throughput scaling factor for PU2RC is
upper bounded as
\begin{equation}\label{Eq:ThputUB:NoSNR}
\lim_{U\rightarrow \infty}    \frac{R}{N_t\log\log U} \leq 1.
\end{equation} }
\end{lemma}
\begin{proof}
See Appendix~\ref{App:ThputUB:NorSNR}.\end{proof}

Next, the upper-bound in \eqref{Eq:ThputUB:NoSNR} is shown to be
achievable. Thereby,  the throughput scaling law of PU2RC in the
normal SNR regime is obtained as shown in the following proposition.
\begin{proposition}\label{Prop:Thput:NorSNR}\emph{
In the normal SNR regime, the throughput scaling law for PU2RC is
\begin{equation}\label{Eq:ScalLaw:NorSNR}
\lim_{U\rightarrow\infty}\frac{R}{N_t\log\log U}=1.
\end{equation}\vspace{0pt}
}
\end{proposition}
\vspace{-10pt}\begin{proof}See Appendix~\ref{App:Thput:NorSNR}.
\end{proof}
The proof uses Lemma~\ref{Lem:LargeNum} on the uniform
convergence in the weak law of the large number. As shown in the proof,
to achieve the throughput scaling law in \eqref{Eq:ScalLaw:NorSNR}, the quantization errors and channel power of scheduled users must scale with the number of users $U$
as $\frac{1}{\log U}$ and $\log U$, respectively. This suggests that
a scheduler for the normal SNR regime should schedule users with both small quantization errors and large channel power as $U$ increases.

\subsection{Interference-Limited
Regime}\label{Section:InterfRegime} In this section, the throughput
scaling  law of PU2RC is analyzed for the interference-limited  regime where
interference dominates over noise. By omitting the noise term, the SINR in \eqref{Eq:SINR:Exp} for the interference-limited regime reduces to
\begin{equation}\label{Eq:SINR:HighSNR}
\SINR^{(\alpha)}_u = \frac{1}{\epsilon_u}-1
\end{equation}
where the superscript $(\alpha)$ identifies the interference-limited
regime. By substituting \eqref{Eq:SINR:HighSNR} into
\eqref{Eq:Thput:Exp}, the throughput for the interference-limited
regime is written as
\begin{equation}
R^{(\alpha)} = E\left[\max_{1\leq m\leq
M}\sum_{n=1}^{N_t}\log\left(\max_{u\in \mathcal{I}^{(m)}_{n}
}\frac{1}{\epsilon_u}\right)\right]. \label{Eq:Thput:HiSNR}
\end{equation}
The scaling law of $R^{(\alpha)}$ with $U$ is obtained as follows.

The upper-bound of the scaling factor of $R^{(\alpha)}$ with $U$ is shown
in the following lemma.
\begin{lemma}\label{Lem:ThputUB:HiSNR} \emph{In the interference limited regime, the
throughput scaling factor is upper bounded as
\begin{equation}\label{Eq:ThputUB:HiSNR}
\lim_{U\rightarrow\infty}\frac{R^{(\alpha)}}{\frac{N_t}{N_t-1}\log
U}\leq 1.\vspace{0pt}
\end{equation}}
\end{lemma}
\vspace{10pt} \begin{proof} See
Appendix~\ref{App:ThputUB:HiSNR}.\end{proof} This proof uses
Lemma~\ref{Lem:LogEps} in Section~\ref{Section:QuantErr}.

Next, the equality in \eqref{Eq:ThputUB:HiSNR} is shown to be
achievable. The main result of this section is summarized in the
following proposition.
\begin{proposition}\label{Prop:Thput:HighSNR}
In the interference-limited regime, the throughput scaling law for
PU2RC is
\begin{equation}\label{Eq:ScalLaw:HiSNR}
\lim_{U\rightarrow\infty}\frac{R^{(\alpha)}}{\frac{N_t}{N_t-1}\log U}=1.
\end{equation}
\vspace{0pt}
\end{proposition}
\vspace{-10pt} \begin{proof} See
Appendix~\ref{App:Thput:HighSNR}.\end{proof} Again, this proof makes
use of Lemma~\ref{Lem:LargeNum} on the uniform convergence in the
weak law of large numbers. By comparing Propositions~\ref{Prop:Thput:NorSNR} and \ref{Prop:Thput:HighSNR}, the throughput scales as $\frac{N_t}{N_t-1}\log U$ in the interference-limited regime but $N_t\log\log U$ otherwise. The reason for this difference is that the asymptotic throughput is determined by the channel power ($\rho$) in the normal SNR and noise-limited regimes, but by the CSI quantization errors ($\epsilon$) of scheduled users in the interference-limited regime. In the normal SNR and noise-limited regimes, the asymptotic throughout can be written as $N_t\E[\log\rho]$,  where $\rho$ scales as $\log U$ due to multiuser diversity gain. In the interference-limited regime, the asymptotic throughput is given as $N_t\E[-\log\epsilon]$ and the scaling law of $\epsilon$ is $U^{-\frac{1}{N_t-1}}$.

A few remarks are in order.
\begin{enumerate}

\item The linear scaling factor in \eqref{Eq:ScalLaw:HiSNR}, namely
$N_t/(N_t-1)$, is smaller than $N_t$, which is the number of
available spatial degrees of freedoms. This indicates the loss in
multiplexing gain for $N_t\geq 3$ in the interference-limited
regime. Such loss is not observed in the normal SNR  (cf.
Proposition~\ref{Prop:Thput:NorSNR}) or noise-limited (cf.
Proposition~\ref{Prop:Thput:LoSNR}) regimes.

\item In the interference-limited regime, scheduling users with
small channel-shape quantization errors is sufficient for ensuring
optimal throughput scaling. The reason is that the SINR in
\eqref{Eq:SINR:HighSNR} depends only on the quantization error.

\item In the interference-limited regime, the throughput scaling law for PU2RC is identical to that for ZF-SDMA
\cite[Theorem~2]{YooJindal:FiniteRateBroadcastMUDiv:2007}\footnote{Note that \cite[(45)]{YooJindal:FiniteRateBroadcastMUDiv:2007} gives the throughput scaling law for a single scheduled user. Multiplication of this result with $N_t$ gives the identical throughput scaling law for PU2RC as shown in Proposition~\ref{Prop:Thput:HighSNR}.}.
\end{enumerate}

\subsection{Noise-Limited Regime}\label{Section:NoiseRegime}
In this section, the throughput scaling  law of PU2RC in the
noise-limited regime is analyzed, where noise dominates over multiuser interference. By removing the
interference term $(\gamma\rho_u\epsilon_u)$ in \eqref{Eq:SINR:Exp},
the SINR for the noise-limited regime is given as
\begin{equation}\label{Eq:SINR:LoSNR}
\SINR^{(\beta)}_u = \gamma\rho_u(1-\epsilon_u)
\end{equation}
where the superscript ${(\beta)}$ specifies the noise-limited
regime. By substituting \eqref{Eq:SINR:LoSNR} into
\eqref{Eq:Thput:Exp}, the corresponding throughput is written as
\begin{equation}
R^{(\beta)} = E\left\{\max_{1\leq m\leq
M}\sum_{n=1}^{N_t}\log\left[1+\max_{u\in \mathcal{I}^{(m)}_{n}
}\gamma\rho_u(1-\epsilon_u)\right]\right\}. \label{Eq:Thput:LoSNR}
\end{equation}

The scaling law of $R^{(\beta)}$ with $U$ for $U\rightarrow\infty$ is
obtained as shown in the following proposition.
\begin{proposition}\label{Prop:Thput:LoSNR}\emph{
In the noise-limited regime, the throughput for PU2RC  scales as
follows
\begin{equation}\label{Eq:ScalLaw:LoSNR}
\lim_{U\rightarrow\infty}\frac{R^{(\beta)}}{N_t\log\log U}=1.
\end{equation}\vspace{0pt}
}
\end{proposition}\vspace{-15pt}
\begin{proof} See Appendix~\ref{App:Thput:LoSNR}. \end{proof} By
comparing Proposition~\ref{Prop:Thput:NorSNR} and
\ref{Prop:Thput:LoSNR}, the throughput scaling laws are observed to
be identical for both the normal SNR  and noise-limited regimes.
Moreover, as reflected in the proof, to achieve the optimal
throughput scaling law, scheduled users in the noise-limited regime
are required to have channel power scaling as $\log U$ and
quantization errors smaller than a constant $d_{\min}$ defined in
\eqref{Eq:dmin}. Thus, for the noise-limited regime, channel power
is a more important scheduling criterion than quantization errors.

\subsection{Non-Asymptotic Regimes}\label{Section:NonAsympRegime}
In preceding sections, the throughput scaling laws for PU2RC are
derived for different asymptotic regimes characterized by an
asymptotically large number of users ($U\rightarrow\infty$). In this
section, these asymptotic scaling laws are compared with their
counterparts in the non-asymptotic regimes corresponding to a finite
number of users ($U< \infty$). The purpose of such a comparison is
to evaluate the usefulness of the asymptotic results derived in
previous section for characterizing the throughput of practical
PU2RC systems.

\begin{figure}
\centering
  \includegraphics[width=10cm]{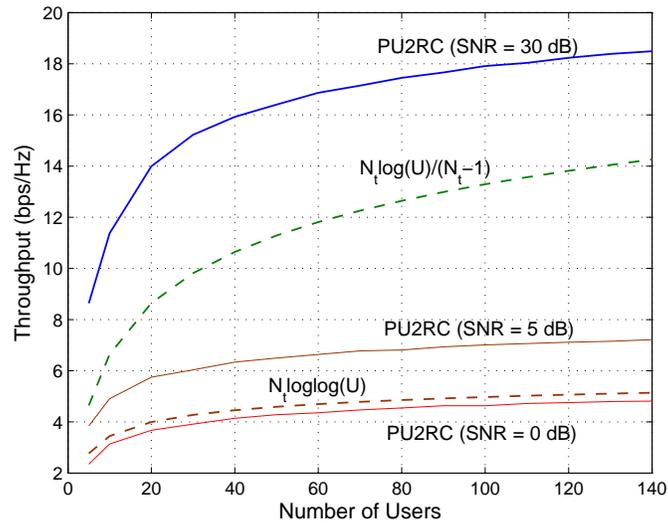}\\
  \caption{Comparison between asymptotic and non-asymptotic throughput scaling laws for PU2RC
  for  $\SNR = \{0, 5, 30\}$~dB, the codebook size $N=16$,  and the number of transmit antennas $N_t=2$.}\label{Fig:ThputScale}
\end{figure}

For this purpose, Fig.~\ref{Fig:ThputScale} shows the throughput
versus number of users curves for the SNR values of $\{0, 5, 30\}$~dB, corresponding respectively to the noise-limited, the normal SNR
and the interference-limited regimes. The range of the number of
users is $1\leq U \leq 140$, the number of transmit antennas is
$N_t=2$ and the codebook size is $N=16$. The above curves present
the PU2RC throughput scaling laws in the non-asymptotic regimes.
Also plotted in Fig.~\ref{Fig:ThputScale}  are the curves defined by
the asymptotic throughput scaling law $\frac{N_t}{N_t-1}\log U$ for
the interference-limited regime (cf.
Proposition~\ref{Prop:Thput:HighSNR}) and $N_t\log\log U$ for both
the normal SNR  and the noise-limited regimes (cf.
Proposition~\ref{Prop:Thput:NorSNR} and \ref{Prop:Thput:LoSNR}). As
observed from Fig.~\ref{Fig:ThputScale},  as the number of users
increases, the non-asymptotic curve for SNR = 30~dB becomes parallel
to the curve following the asymptotic throughput scaling law
$\frac{N_t}{N_t-1}\log U$. Likewise, the non-asymptotic curves for
SNR = 0~dB and 5~dB have the same slopes as the corresponding
asymptotic curve defined by $N_t\log\log U$. Therefore, the
asymptotic throughput scaling laws also hold in the non-asymptotic
regimes. Note that the gaps between the asymptotic and
non-asymptotic curves are throughput constant factors that become
insignificant in the asymptotic regimes ($U\rightarrow\infty$).

\section{Numerical Results}\aftersect\label{Section:Numerical} In this
section, various numerical results are presented. In
Section~\ref{Section:Num:ShFb}, the effect of increasing channel
shape feedback on throughput is investigated. In
Section~\ref{Section:Num:ZFSDMA}, for an increasing number of users,
the throughput  of PU2RC is evaluated against  that of ZF-SDMA in
\cite{YooJindal:FiniteRateBroadcastMUDiv:2007} as well as the upper
bound achieved by dirty paper coding (DPC) and multiuser water
filling \cite{Jindal:IterativeWaterFilling:2005}. For simplicity,
Assumption~\ref{AS:ChanGain} is made and thus the SINR feedback is assumed
perfect for all algorithms in comparison. In
Section~\ref{Sectioin:Num:SINRQuant}, the capacity loss due to the
SINR quantization is characterized.

\subsection{Effect of Increasing Channel Shape Feedback} \label{Section:Num:ShFb}
\begin{figure}
\centering \subfigure[Small numbers of users]{\hspace{-20pt}\includegraphics[width=9cm]{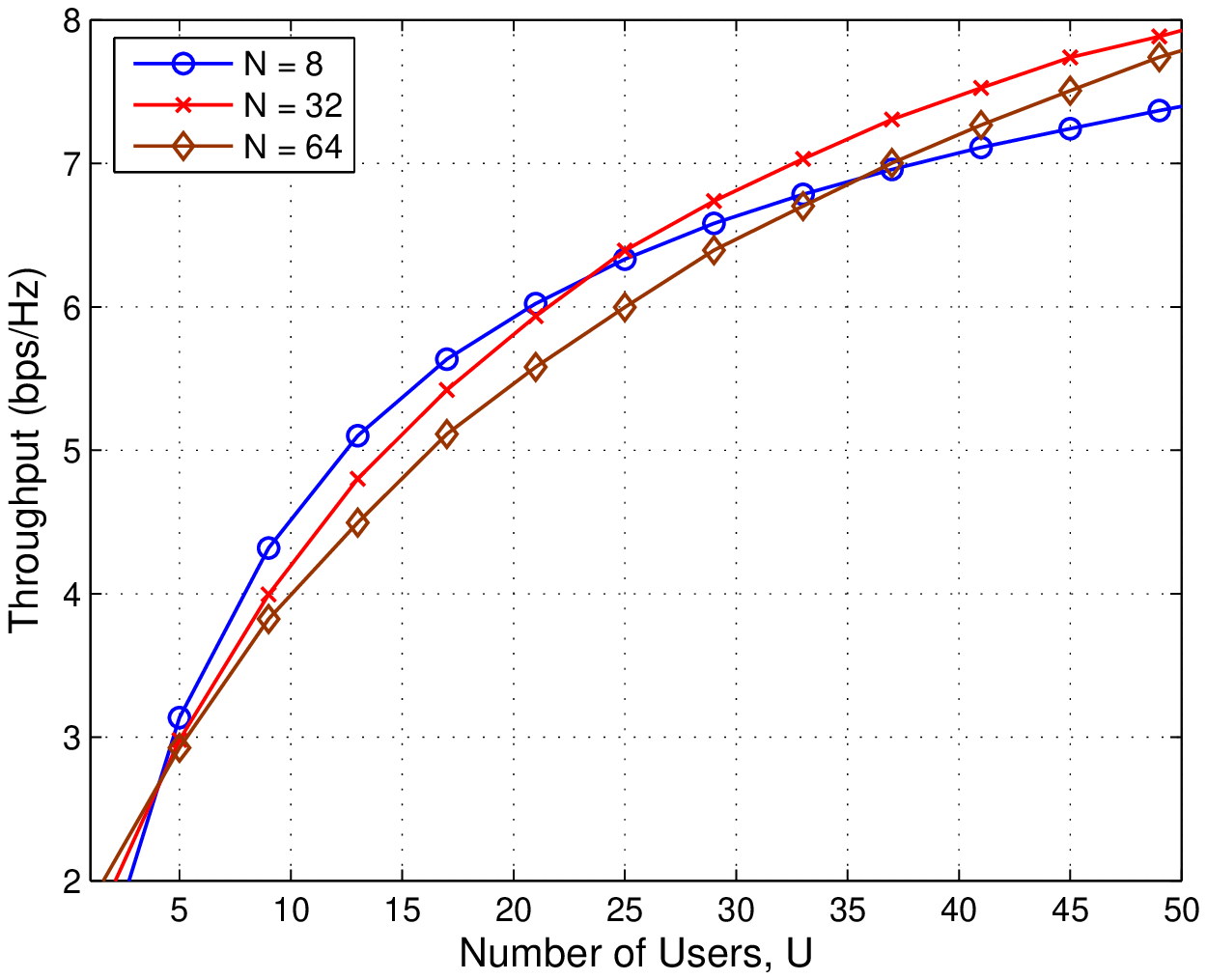}}\hspace{-20pt}
\subfigure[Large numbers of users]{\includegraphics[width=9cm]{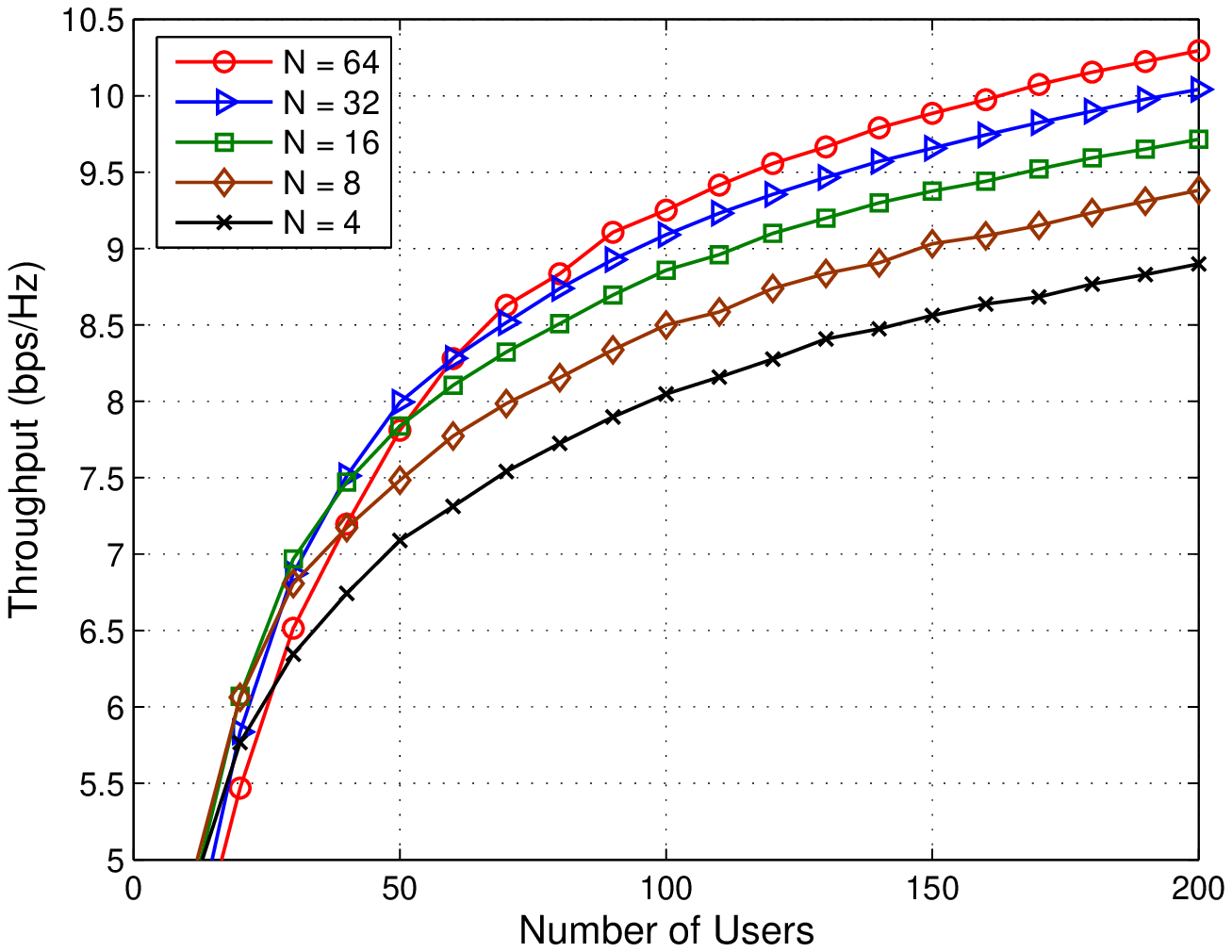}\hspace{-20pt}}\\
  \caption{Throughput of  PU2RC for an increasing number of users $U$,
  $\SNR = 5$~dB, and the number of transmit antennas $N_t=4$.}\label{Fig:PU2RC}
\end{figure}

For PU2RC,
increasing channel shape feedback does not necessarily lead to higher
throughput as shown in Fig.~\ref{Fig:PU2RC}. In
Fig.~\ref{Fig:PU2RC}, the curves of PU2RC throughput versus the
number of users $U$ are plotted for different codebook sizes $N$.
The SNR is 5~dB and the number of transmit antennas is $N_t=4$.
Fig.~\ref{Fig:PU2RC}(a) and Fig.~\ref{Fig:PU2RC}(b) show the small
($1\leq U \leq 50$)  and the large user ranges ($1\leq U\leq 200$),
respectively. As observed from Fig.~\ref{Fig:PU2RC}(a), in the range
of $4\leq U \leq 22$, increasing $N$ decreases the throughput. The
reason is that a larger codebook size divides the user pool because
each user is associated with only one codebook vector (cf.
Section~\ref{Section:BeamSel}). Consequently, increasing the
codebook size reduces the probability of finding scheduled users
with large channel gains and also associated with the same
orthonormal basis in the codebook. Nevertheless, such an adverse
effect of increasing the codebook size diminishes as the number of
users increases.  As shown in Fig.~\ref{Fig:PU2RC}, for $U\geq 70$,
a larger codebook size results in higher throughput. The above
results motivate the need for choosing an optimal codebook size for
a given number of users.

\subsection{Comparison with ZF-SDMA and Dirty Paper Coding}\label{Section:Num:ZFSDMA} Presently, PU2RC and ZF-SDMA
\cite{YooJindal:FiniteRateBroadcastMUDiv:2007,LTEZFMIMO:06,LTEMIMO:06}
are two main solutions for multiuser MIMO downlink for 3GPP-LTE. In
this section, their performance is compared using numerical results.
Moreover, the throughput of PU2RC is evaluated against the
upper-bound achieved by dirty paper coding.

\begin{figure}
\centering \subfigure[Small Numbers of Users]{\hspace{-20pt}
\includegraphics[width=9cm]{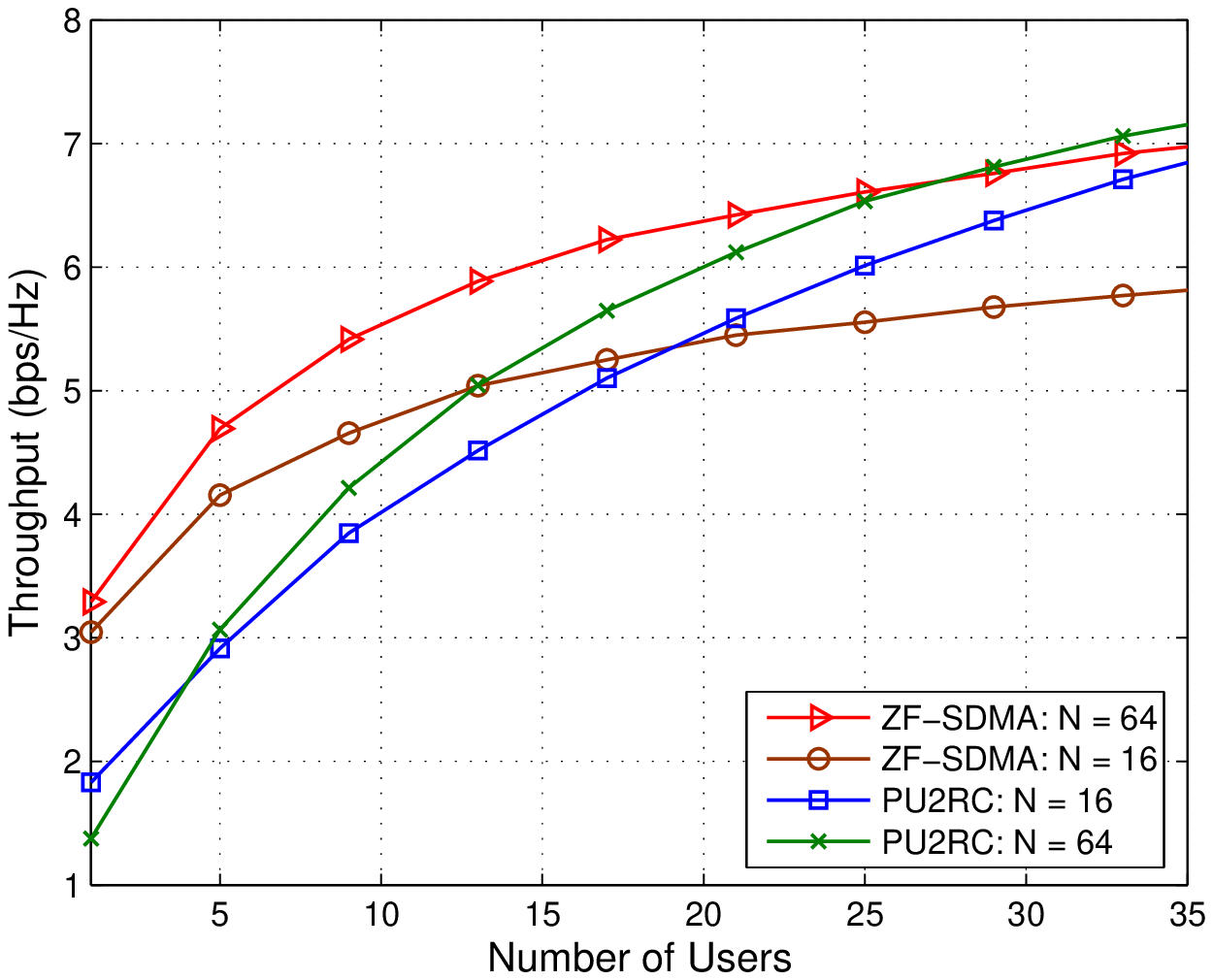}}\hspace{-20pt}
\subfigure[Large Numbers of Users]{\includegraphics[width=9cm]{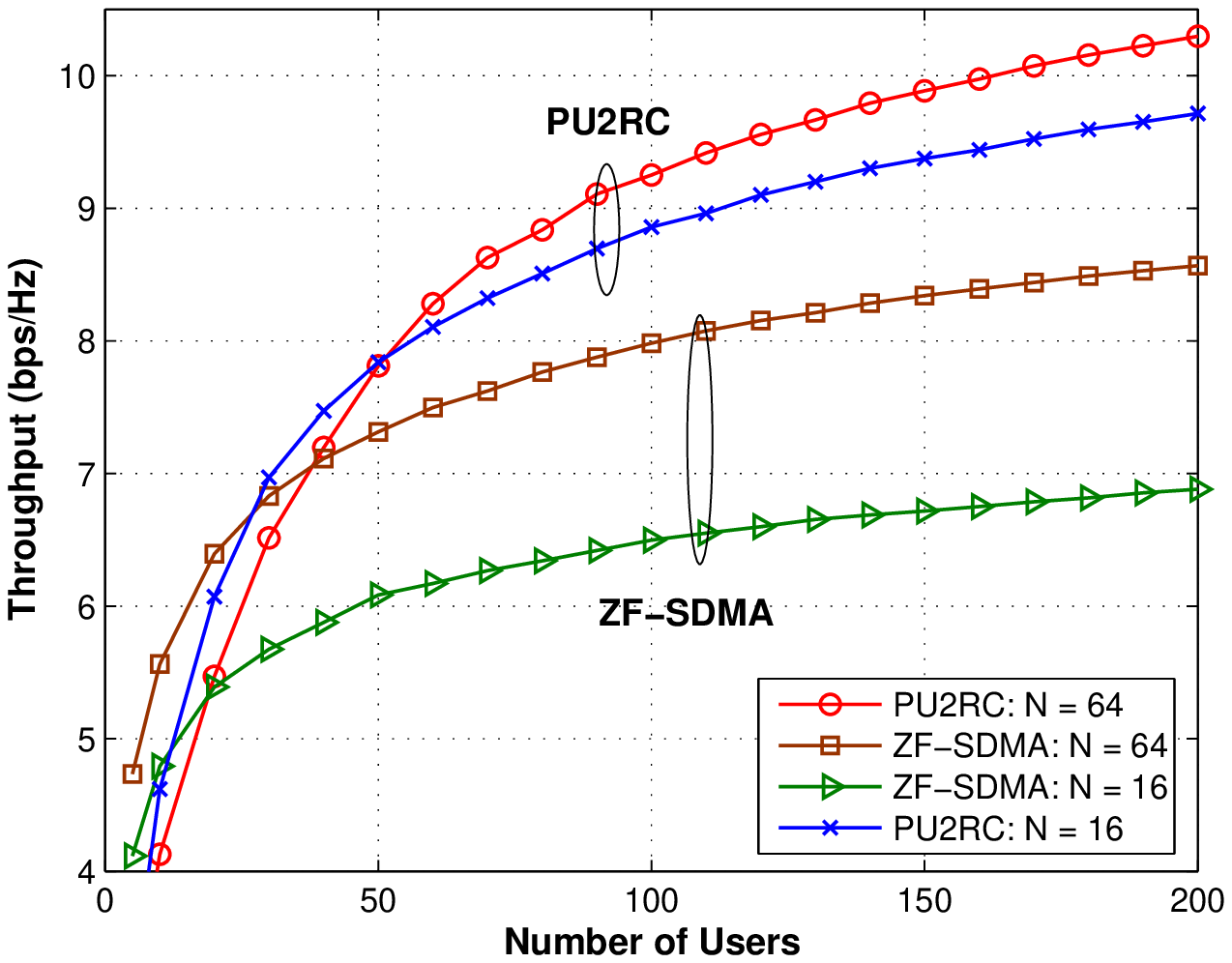}}\hspace{-20pt}\\
 %   \subfigure[Two Transmit Antennas ($N_t=2$)]{\includegraphics[width=11cm]{CmpYoo_2Ant.eps}}\\
  \caption{Throughput  comparison between PU2RC and ZF-SDMA for an increasing number of users $U$,
  $\SNR = 5$~dB, and the number of transmit antennas $N_t=4$.}\label{Fig:CmpYoo}
\end{figure}

In Fig.~\ref{Fig:CmpYoo}, the throughput  of PU2RC is compared with
that of ZF-SDMA for an increasing number of users. The number of
transmit antenna is $N_t=4$ and the SNR is 5 dB. Moreover, the
codebook sizes $N=\{4, 8, 16, 32\}$ for channel shape quantization
are considered. As in
\cite{YooJindal:FiniteRateBroadcastMUDiv:2007},
 the threshold $0.25$ is applied in the greedy-search scheduling for ZF-SDMA.
Fig.~\ref{Fig:CmpYoo}(a) and Fig.~\ref{Fig:CmpYoo}(b) show
respectively the small ($1\leq U \leq 35$) and the large  ($1\leq U
\leq 200$) user ranges. As observed from Fig.~\ref{Fig:CmpYoo}(a),
for a given codebook size (either $N=16$ or $N=64)$, PU2RC achieves
higher throughput than ZF-SDMA for a relative large number of users
but the reverse holds for a smaller user pool. Specifically, in
Fig.~\ref{Fig:CmpYoo}(a),  the throughput curves for PU2RC and
ZF-SDMA cross at $U=19$ for $N=16$ and at $U=27$ for $N=64$. For a
sufficiently large number of users, PU2RC always outperforms
ZF-SDMA in terms of throughput as shown in Fig.~\ref{Fig:CmpYoo}(b).
Furthermore, compared with ZF-SDMA, PU2RC is found to be more robust
against CSI quantization errors. For example, as observed from
Fig.~\ref{Fig:CmpYoo}(b), for $U=100$, the throughput loss for PU2RC
due to the decrease of the codebook size from $N=64$ to $N=16$ is
0.3~bps/Hz but that for ZF-SDMA is 1.5~bps/Hz.
 The above observations are explained shortly. In
summary, these observations suggest that PU2RC is preferred to
ZF-SDMA for a large user pool but not for a small one.

\begin{figure}
\centering
  \includegraphics[width=10cm]{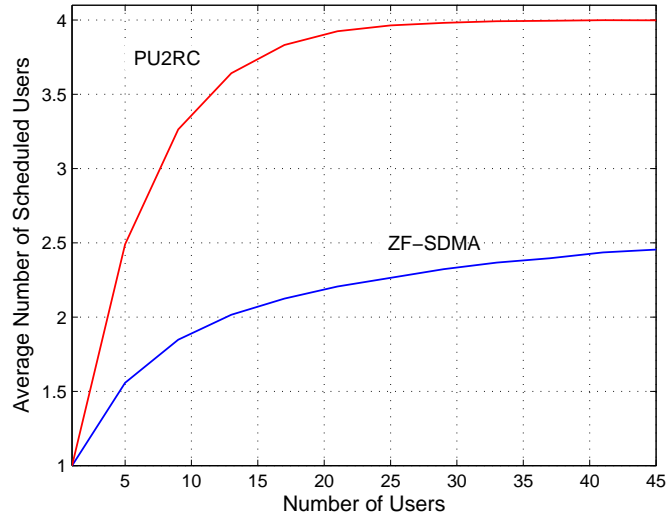}\\
  \caption{The average numbers of scheduled users  for PU2RC and ZF-SDMA for
  $\SNR = 5$~dB, and the number of transmit antennas $N_t=4$.}\label{Fig:NumUsr}
\end{figure}

To explain the observations from Fig.~\ref{Fig:CmpYoo}, the average
numbers of scheduled users for PU2RC and ZF-SDMA are compared in
Fig.~\ref{Fig:NumUsr} for an increasing number of users. It can be
observed from Fig.~\ref{Fig:NumUsr} that PU2RC tends to schedule
more users than ZF-SDMA. First, for a small number of users, interference between
scheduled users can not be effectively suppressed by scheduling, and hence more simultaneous users result in smaller throughput. This explains the observation
from Fig.~\ref{Fig:CmpYoo}(a) that PU2RC achieves lower throughput
than ZF-SDMA due to more scheduled users. Second, for a large user
pool, the channel vectors of scheduled users are close-to-orthogonal
and interference is negligible. Therefore, a larger number of
scheduled users leads to higher throughput. For this reason, PU2RC
outperforms ZF-SDMA for a large number of users as observed from
Fig.~\ref{Fig:CmpYoo}(b). Last, with respect to ZF-SDMA, the better
robustness of PU2RC against CSI quantization errors is mainly due to
the joint beamforming and scheduling (cf.
Section~\ref{Section:BeamSel}). Note that beamforming and scheduling
for ZF-SDMA are performed separately
\cite{YooJindal:FiniteRateBroadcastMUDiv:2007}.

\begin{figure}
\centering
  \includegraphics[width=10cm]{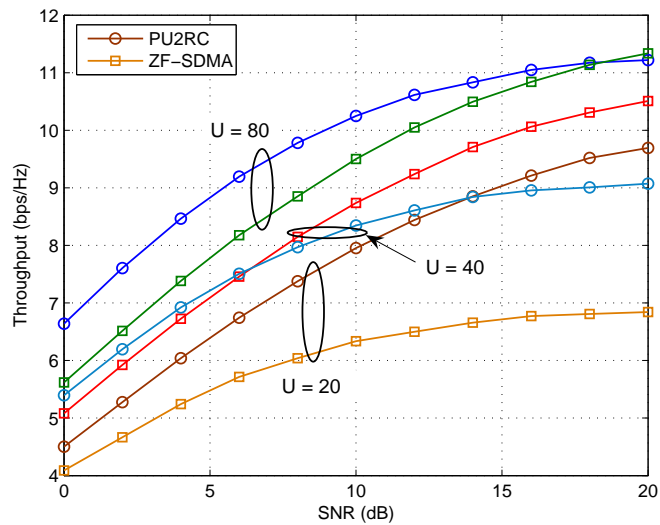}\\
 %   \subfigure[Two Transmit Antennas ($N_t=2$)]{\includegraphics[width=11cm]{CmpYoo_2Ant.eps}}\\
  \caption{Throughput  comparison between PU2RC and ZF-SDMA for an increasing number SNR;
  The codebook size $N=64$ and the number of transmit antennas $N_t=4$.}\label{Fig:CmpYoo_SNR}
\end{figure}

Fig.~\ref{Fig:CmpYoo_SNR} compares the throughput of PU2RC and
ZF-SDMA for an increasing SNR. The number of transmit antennas is  $N_t=4$ and the
codebook size is $N=64$. As observed from Fig.~\ref{Fig:CmpYoo_SNR},
for the number of users $U=20$, PU2RC achieves lower throughput than
ZF-SDMA over the range of SNR under consideration ($0\leq
\text{SNR}\leq 20$ dB). Nevertheless, for larger numbers of users
($U=40$ or 80), PU2RC outperforms ZF-SDMA for a subset of the SNRs.
Specifically, the throughput versus SNR curves for PU2RC and ZF-SDMA
crosses at SNR=7~dB for $U=40$ and at SNR=18~dB for $U=80$. The
above results suggest that in the practical range of SNR, PU2RC is
preferred to ZF-SDMA only if the user pool is sufficiently large.

\begin{figure}
\centering
\includegraphics[width=11cm]{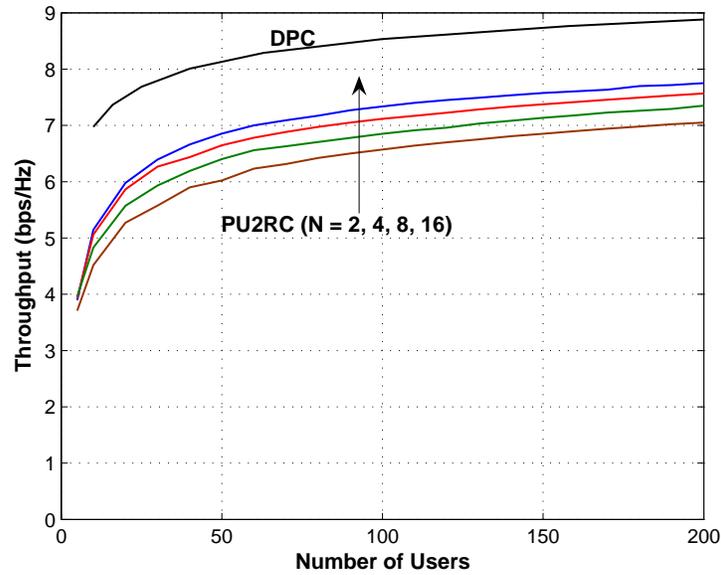}\\
  \caption{Comparison between the throughput  of PU2RC and its upper bound
  achieved by dirty paper coding (DPC) and multiuser iterative water-filling for an increasing number of users $U$,
  $\SNR = 5$~dB, and the number of transmit antennas $N_t=2$.}\label{Fig:CmpDPC}
\end{figure}

Fig.~\ref{Fig:CmpDPC} compares the throughput  of PU2RC with an upper
bound achieved by \emph{dirty paper coding} (DPC) and multiuser
iterative water-filling \cite{Jindal:IterativeWaterFilling:2005}. A
smaller number of antennas $N_t=2$ is chosen to reduce the high
computational complexity of iterative water-filling for a large
number of users. Hence,  each user has a $2\times 1$ multiple-input-single-output (MISO) channel. Moreover,
$\SNR = 5$ dB and the channel shape codebook size is $N=\{2, 4, 8,
16\}$. As observed from Fig.~\ref{Fig:CmpDPC}, the gap between the
throughput  of PU2RC and its upper bound narrows as the number of
users $U$ or the codebook size $N$ increases. At $U=200$ and $N=16$,
PU2RC achieves about $85\%$ of the sum capacity of DPC.

\subsection{Effect of SINR Quantization}\label{Sectioin:Num:SINRQuant}
In this section, using numerical results, a small number of bits for
SINR feedback is found sufficient for making the capacity loss due
to SINR quantization negligible.

\begin{figure}
\centering
\includegraphics[width=11cm]{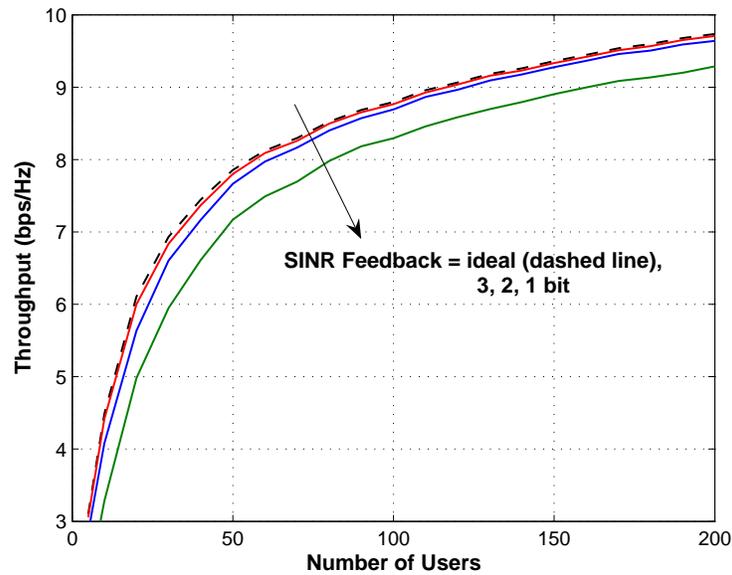}\\
  \caption{The effect of SINR quantization for $\SNR = 5$~dB, the number of transmit antennas $N_t=4$, and
  the codebook size for channel shape quantization is $N=16$.}\label{Fig:CmpQSINR}
\end{figure}

For PU2RC, Fig.~\ref{Fig:CmpQSINR} compares the cases of perfect and
quantized SINR feedback. For quantizing SINR, a scalar quantizer
using a squared-error distortion function is employed \cite{GerGra:VectQuanSignComp:92}.
Moreover, the
quantizer has a simple codebook containing evenly spaced scalars in
the SINR range corresponding to a probability of $99\%$. The number
of transmit antenna is $N_t=4$, the SINR is 5 dB and the codebook
size for channel shape quantization is $N=16$. As observed from
Fig.~\ref{Fig:CmpQSINR}, 2 bits of SINR feedback per user causes
only marginal loss in throughput  with respect to the perfect SINR
feedback. Such loss is negligible for 3-bit feedback. Therefore, a
few bits of SINR feedback from each user is almost as good as the perfect
case, which justifies Assumption~\ref{AS:ChanGain}.

\section{Conclusion}\aftersect\label{Section:Conclusion}
This paper presents asymptotic throughput scaling laws for SDMA with
orthogonal beamforming known as PU2RC for different SNR regimes. In the
interference limited regime, the throughput of PU2RC is shown to
scale logarithmically with the number of users but does not increase
with the number of antennas. In the normal SNR  or noise-limited regimes,
the throughput of PU2RC is found to scale double logarithmically
with the number of users and linearly with the number of antennas at
the base station. Numerical results showed that PU2RC can achieve
significant gains in throughput  with respect to ZF-SDMA for the
same amount of CSI feedback.

This paper focuses on the scheduling criterion of maximizing
throughput. The design and performance analysis of PU2RC based on
the criterion of proportional fairness is a topic under
investigation. Furthermore, the optimal deign for PU2RC for the
non-asymptotic regime of the user pool remains as an open issue.

\appendices
\renewcommand{\baselinestretch}{1.5}\normalsize

\section{Proof of Lemma~\ref{Lem:LargeNum}}\label{App:LargeNum}
Lemma~4.8 in \cite{GuptaKumar:CapWlssNetwk:2000} can be generalized
from $\mathds{R}^3$ to $\mathds{C}^{N_t}$ as follows.
\cite[Lemma~4.8]{GuptaKumar:CapWlssNetwk:2000} concerns
$N$ congruent disks on the surface of a sphere in $\mathds{R}^3$, and its
derivation relies on two results: the first one is  the \emph{stereographic
projection} \cite{BrannanBook:Geometry:99} that one-to-one maps a
point on the surface of the sphere to a point on a plane both in
$\mathds{R}^3$; the second is that the Vapnik–-Chervonenkis dimension of a set of disks on a plane
in $\mathds{R}^3$ is three \cite{GuptaKumar:CapWlssNetwk:2000}.
A unit hyper-sphere in $\mathds{C}^{N_t}$ can be treated as one in $\mathds{R}^{2N_t}$\cite{MukSabETAL:BeamFiniRateFeed:Oct:03}. Thereby, the \emph{stereographic
projection} also exists between a unit hyper-sphere and a hyper-plane in $\mathds{C}^{N_t}$ \cite{RosenfeldBook:GemoetryLieGroups}. Next, following the same procedure as in  \cite[Lemma~4.6]{GuptaKumar:CapWlssNetwk:2000}, the Vapnik–-Chervonenkis dimension of a set of disks on a hyper-plane in  $\mathds{C}^{N_t}$ is shown to be also three. Based on the two results obtained above for $\mathds{C}^{N_t}$, the remaining steps for proving Lemma~\ref{Lem:LargeNum} are identical to those for \cite[Lemma~4.8]{GuptaKumar:CapWlssNetwk:2000} and are thus omitted.

\section{Proof of Lemma~\ref{Lem:EpsCDF}}\label{App:EpsCDF}

Since the orthonormal bases in the codebook $\mathcal{F}$ are
independently and randomly generated, the complementary CDF
\eqref{Eq:CDF:Def} can be equivalently expressed as
\begin{equation}\label{Eq:CDF:c}
\Pr(\epsilon\geq \delta) = \prod_{m=1}^M\Pr\left(\bs\notin
\cup_{\bv\in\mathcal{V}^{(m)}}B_{\delta}(\bv)\right),
\end{equation}
where $\mathcal{V}^{(m)}$ denotes the $m$th orthonormal basis in
$\mathcal{F}$. Given that $\bs$ is isotropically distributed on the
unit sphere, \eqref{Eq:CDF:c} can be re-written in terms of the
\emph{volume} of sphere caps \cite{GerGra:VectQuanSignComp:92}
\begin{equation}
\Pr(\epsilon \geq \delta) = \prod_{m=1}^M\left\{1-
\vol[\cup_{\bv\in\mathcal{V}^{(m)}}B_{\delta}(\bv)]\right\}.
\label{Eq:CDF:d}
\end{equation}
Since the sphere caps $\{B_{\delta}(\bv)\}\mid
\bv\in\mathcal{V}^{(m)}\}$ are non-overlapping for $\delta \leq
\tfrac{1}{2}$ and the volume of each sphere cap is
$\vol[B_{\delta}(\bv)]=\delta^{N_t-1}$ as obtained in
\cite{SharifHassibi:CapMIMOBroadcastPartSideInfo:Feb:05}, we can
obtain from \eqref{Eq:CDF:d}
\begin{equation}
\Pr(\epsilon \geq \delta) = \prod_{m=1}^M\left(1-N_t
\delta^{N_t-1}\right), \quad 0\leq \delta \leq \tfrac{1}{2}.
\label{Eq:CDF:e}
\end{equation}
The desired result in \eqref{Eq:EpsCDF} follows from the last
equation. Moreover, from \eqref{Eq:CDF:d} and for
$\bv\in\mathcal{F}$,
 $\Pr(\epsilon \geq \delta) \leq \prod_{m=1}^M\left\{1-
\vol[B_{\delta}(\bv)]\right\} = \prod_{m=1}^M\left\{1-
\delta^{N_t-1}\right\}$, which gives the inequality in the lemma.

%-------------- Lemma: Log Epsilon ----------------------
\section{Proof of Lemma~\ref{Lem:LogEps}}\label{App:LogEps}
The minimum of $M$ i.i.d. Beta $(N_t, 1)$ random variables, denoted
as $\{\beta_1, \beta_2, \cdots, \beta_M\}$, has the following CDF
\cite{Jindal:MIMOBroadcastFiniteRateFeedback:06}
\begin{equation}\label{Eq:BetaCDF}
\Pr\left(\min_{1\leq m\leq M}\beta_m\geq b \right) =(1-b^{N_t-1})^M.
\end{equation}
From \eqref{Eq:BetaCDF} and Lemma~\ref{Lem:EpsCDF}
\begin{equation}
N_t^{\frac{1}{N_t-1}}\epsilon \cong \min_{1\leq m\leq M}\beta_m, \quad
\epsilon \leq \frac{1}{2}
\end{equation}
where $\cong$ represents equivalence in distribution. The above
equivalence results in the following equality $(a)$
\begin{eqnarray}
\E\left[-\log\left(N_t^{\frac{1}{N_t-1}}\epsilon\right)\right]&\leq
&\E\left[-\log\left(N_t^{\frac{1}{N_t-1}}\epsilon\right)\mid 0\leq
\epsilon\leq \frac{1}{2}\right] \nn\\
&\overset{(a)}{=}& \E\left[-\log\left(\min_{1\leq m\leq
M}\beta_m\right)\mid 0\leq
\min_{1\leq m\leq M}\beta_m\leq \frac{1}{2}N_t^{\frac{1}{N_t-1}}\right] \nn\\
&\leq&
\frac{\E\left[-\log\left(\min_{m}\beta_m\right)\right]}{\Pr\left(0\leq
\min_{m}\beta_m\leq \frac{1}{2}N_t^{\frac{1}{N_t-1}}\right)}.
\label{Eq:App:a}
\end{eqnarray}
 As shown in \cite[Lemma~4]{Jindal:MIMOBroadcastFiniteRateFeedback:06}
\begin{equation}\label{Eq:App:b}
\frac{\log M}{N_t-1} \leq
\E\left[-\log\left(\min_{m}\beta_m\right)\right]\leq \frac{\log M +
1}{N_t-1}.
\end{equation}
By combining \eqref{Eq:BetaCDF}, \eqref{Eq:App:a}, and \eqref{Eq:App:b}, the desired inequality follows.

\section{Proof of Lemma~\ref{Lem:ThputUB:NoSNR}}\label{App:ThputUB:NorSNR}
From \eqref{Eq:SINR:Exp} and \eqref{Eq:Thput:Exp}
\begin{eqnarray}
R &=& E\left[\max_{1\leq m\leq
M}\sum_{n=1}^{N_t}\log\left(\max_{u\in \mathcal{I}^{(m)}_{n}
}\frac{1+\gamma\rho_u}{1+\gamma\rho_u\epsilon_u}\right)\right]\nn\\
&\leq& E\left[\max_{1\leq m\leq
M}\sum_{n=1}^{N_t}\log\left(1+\gamma\max_{u\in \mathcal{I}^{(m)}_{n}
}\rho_u\right)\right]\label{Eq:App:k}\\
&\leq& E\left[\sum_{n=1}^{N_t}\log\left(1+\gamma\max_{1\leq m\leq
M}\max_{u\in \mathcal{I}^{(m)}_{n} }\rho_u\right)\right]\nn\\
&=& N_tE\left[\log\left(1+\gamma\max_{1\leq u\leq
U}\rho_u\right)\right].\label{Eq:App:g}
\end{eqnarray}
The following result is well-known from extreme value theory (see
e.g.
\cite[(A10)]{SharifHassibi:CapMIMOBroadcastPartSideInfo:Feb:05})
\begin{equation}\label{Eq:ExtValTheo}
\Pr\left(\left|\max\limits_{1\leq u\leq U}\rho_{u}-\log U\right|<
O(\log\log U)\right) > 1 - O\left(\frac{1}{\log U}\right).
\end{equation}
From \eqref{Eq:ExtValTheo} and \eqref{Eq:App:g}
\begin{eqnarray}\label{Eq:App:A} R
&\leq& N_tE\left\{\log\left[1+\gamma\log U -\gamma O(\log\log
U)\right]\right\}\Pr\left(\max_{1\leq u\leq U}\rho_u \leq \log U -
O(\log\log U)\right) + \nn\\
&& N_tE\left[\log\left(1+\gamma\sum_{u=1}^U\rho_u\right)\right]O\left(\frac{1}{\log U}\right)\nn\\
&\overset{(a)}{\leq}& N_tE\left\{\log\left[1+\gamma\log U -\gamma
O(\log\log U)\right]\right\} + N_t\log\left(1+\gamma N_t U
\right)O\left(\frac{1}{\log U}\right)
\end{eqnarray}
where  $(a)$ is obtained by using Jensen's inequality. The desired
inequality follows from \eqref{Eq:App:A}.

\section{Proof of Proposition~\ref{Prop:Thput:NorSNR}}\label{App:Thput:NorSNR}
Define a set of disks on the unit hyper-sphere as
\begin{equation}\label{Eq:SphCap:Def}
\mathcal{B}^{(m)}_n(d) = \left\{\bs\in \mathds{C}^{N_t}\mid
\|\bs\|^2=1,  \ 1-|\bs^\dagger\bv^{(m)}_n|^2\leq d \right\} \quad
1\leq m \leq M, 1\leq n\leq N_t
\end{equation}
where $d$ is the radius of $\mathcal{B}^{(m)}_n(d)$. Furthermore,
define the user index sets
\begin{equation}\label{Eq:IndexSet:T:Hat}
\hat{\mathcal{T}}^{(m)}_n = \left\{1\leq u \leq U\mid \bs_u \in
\mathcal{B}^{(m)}_n\left(\frac{1}{\log U}\right)\right\} \quad 1\leq
m\leq M, 1\leq n\leq N_t
\end{equation}
where the disk $\mathcal{B}^{(m)}_n$ is defined in
\eqref{Eq:SphCap:Def}. By applying Lemma~\ref{Lem:LargeNum} with
$\tau_1=\tau_2=A=\frac{1}{2(\log U)^{N_t-1}}$, we obtain that
\begin{equation}\label{App:LargNumb:Eq1}
\Pr\left(|\hat{\mathcal{T}}^{(m)}_n|\geq \frac{U}{\left(\log
U\right)^{N_t-1}}\right) > 1- \frac{1}{2\left(\log
U\right)^{N_t-1}}\quad \forall \ U> U_o
\end{equation}
where $U_o$ is in \eqref{Eq:Uo}. Let $U_1$ denote a sufficiently
large integer such that $\hat{\mathcal{T}}^{(m)}_{n}\subset
\mathcal{I}^{(m)}_{n}$. From \eqref{Eq:Thput:Exp} and
\eqref{Eq:SINR:Exp} and by replacing $\mathcal{I}^{(m)}_{n}$
with $\hat{\mathcal{T}}^{(m)}_{n}$
\begin{eqnarray}
R &\geq& E\left[\sum_{n=1}^{N_t}\log\left(\max_{u\in
\hat{\mathcal{T}}^{(m)}_{n}
}\frac{1+\gamma\rho_u}{1+\gamma\rho_u\epsilon_u}\right)\right],\quad U\geq U_1\nn\\
&\overset{(a)}{\geq}& E\left[\sum_{n=1}^{N_t}\log\left(\max_{u\in
\hat{\mathcal{T}}^{(m)}_{n}
}\frac{1+\gamma\rho_u}{1+\gamma\rho_u\frac{1}{\log
U}}\right)\right],\quad U\geq U_1\nn\\
&\geq& E\left[\sum_{n=1}^{N_t}\log\left(\frac{1+\gamma\max_{u\in
\hat{\mathcal{T}}^{(m)}_{n} }\rho_u}{1+\frac{\gamma}{\log
U}\max_{u\in \hat{\mathcal{T}}^{(m)}_{n}
}\rho_u}\right)\right],\quad U\geq U_1\label{Eq:App:B}
\end{eqnarray}
where the inequality in (a) holds because
$u\in\hat{\mathcal{T}}^{(m)}_{n} \Rightarrow \epsilon_u\leq
\frac{1}{\log U}$ according to the definition in
\eqref{Eq:IndexSet:T:Hat}. From \eqref{App:LargNumb:Eq1} and
\eqref{Eq:App:B}
\begin{eqnarray}
R &\geq & E\left[\sum_{n=1}^{N_t}\log\left(\frac{1+\gamma\max_{u\in
\hat{\mathcal{T}}^{(m)}_{n} }\rho_u}{1+\frac{\gamma}{\log
U}\max_{u\in \hat{\mathcal{T}}^{(m)}_{n} }\rho_u}\right)\mid
|\hat{\mathcal{T}}^{(m)}_n|\geq \frac{U}{\left(\log
U\right)^{N_t-1}}\right]\times\nn\\
&& \left[1- \frac{1}{2\left(\log U\right)^{N_t-1}}\right]\quad
\forall \ U> \max(U_1, U_o).\nn
\end{eqnarray}
From the last inequality and \eqref{Eq:ExtValTheo},
\begin{eqnarray}
R &\geq& N_t\E\left[
\log\left(1+\frac{\log\tilde{U}-O(\log\log\tilde{U})}{1/\gamma+[\log\tilde{U}+O(\log\log\tilde{U})]\frac{1}{\log
U}}\right)\right]\left(1- \frac{1}{2\left(\log U\right)^{N_t-1}}\right)\times\nn\\
&& \left[1 - O\left(\frac{1}{\log U}\right)\right]^{N_t}, \quad
\forall \ U> \max(U_1, U_o)\nn
\end{eqnarray}
where $\tilde{U}=\frac{U}{(\log U)^{N_t-1}}$. It follows from the
last inequality that
\begin{equation}\label{Eq:ThputLB:NoSNR}
\lim_{U\rightarrow\infty} \frac{R}{N_t\log\log U} \geq 1.
\end{equation}
The desired result is obtained by combining \eqref{Eq:ThputLB:NoSNR}
and Lemma~\ref{Lem:ThputUB:NoSNR}.

\section{Proof of Lemma~\ref{Lem:ThputUB:HiSNR}}\label{App:ThputUB:HiSNR}
From \eqref{Eq:Thput:HiSNR}
\begin{eqnarray}
R^{(a)} &\leq& E\left[\sum_{n=1}^{N_t}-\log\left(\min_{1\leq m\leq
M}\min_{u\in \mathcal{I}^{(m)}_{n} }\epsilon_u\right)\right]\nn\\
&=& N_tE\left[-\log\left(\min_{1\leq u\leq
U}\epsilon_u\right)\right].\label{Eq:App:c}
\end{eqnarray}
In the above equation, $\min_{1\leq u\leq U}\epsilon_u$ follows the
same distribution as the quantization error for an enlarged codebook
having $MU$ orthonormal bases. Therefore, from \eqref{Eq:App:c} and
Lemma~\ref{Lem:LogEps}
\begin{equation}
R^{(a)} \leq     \frac{N_t}{N_t-1}\left\{\frac{\log U +\log M +
1}{1-\left[1-N_t2^{-(N_t-1)}\right]^{MU}} + \log N_t\right\}.
\end{equation}
The desired upper bound of the throughput scaling factor follows
from the last inequality. Note that
$\left\{1-\left[1-N_t2^{-(N_t-1)}\right]^{MU}\right\}\rightarrow 1$
as $U\rightarrow \infty$.

\section{Proof of
Proposition~\ref{Prop:Thput:HighSNR}}\label{App:Thput:HighSNR}
Define the minimum distance of the codebook $\mathcal{F}$ as
\begin{equation}\label{Eq:dmin}
d_{\min} = \min_{\bv,\bv'\in\mathcal{F}} \frac{1 -
|\bv^\dagger\bv'|^2}{4}.
\end{equation} Moreover, similar to \eqref{Eq:IndexSet:T:Hat}, define
the index set of the users in the disk
$\mathcal{B}^{(m)}_n\left(d_{\min}\right)$ (cf.
\eqref{Eq:SphCap:Def}) as
\begin{equation}\label{Eq:IndexSet:T}
\mathcal{T}^{(m)}_n = \left\{1\leq u \leq U\mid \bs_u \in
\mathcal{B}^{(m)}_n\left(d_{\min}\right)\right\}, \quad 1\leq m\leq
M, 1\leq n\leq N_t.
\end{equation}
By the definitions in \eqref{Eq:I-set} and \eqref{Eq:dmin},
$\bs_u\in \mathcal{B}^{(m)}_n\left(d_{\min}\right)\Rightarrow u \in
\mathcal{I}_{m,n}$. Using this fact, a throughput lower bound
follows by replacing $\mathcal{I}_{m,n}$ in \eqref{Eq:Thput:Exp}
with $\mathcal{T}_{m,n}$
\begin{eqnarray}
R^{(a)} &\geq& E\left[\max_{1\leq m\leq
M}\sum_{n=1}^{N_t}\log\left(\max_{u\in \mathcal{T}^{(m)}_{n}
}\frac{1}{\epsilon_u}\right)\right]\nn\\
&\geq& E\left[\sum_{n=1}^{N_t}\log\left(\max_{u\in
\mathcal{T}^{(m)}_{n}
}\frac{1}{\epsilon_u}\right)\right].\label{Eq:App:f}
\end{eqnarray}
By applying Lemma~\ref{Lem:LargeNum} with
$\tau_1=\tau_2=U^{-\frac{1}{2}}$ and $A = d_{\min}^{N_t-1}$, the numbers
of users belonging to the index sets \eqref{Eq:IndexSet:T:Hat}
satisfy
\begin{equation}\label{Eq:App:LargeNumb}
\Pr\left(\min_{m,n}\left|\mathcal{T}^{(m)}_n\right| \geq
d_{\min}^{N_t-1}U-U^{\frac{1}{2}}\right)\geq 1-U^{-\frac{1}{2}}, \quad \ \forall \
U\geq U_o
\end{equation}
where $U_o$ is defined in \eqref{Eq:Uo}. From \eqref{Eq:App:f} and
\eqref{Eq:App:LargeNumb}
\begin{equation}
R^{(a)} \geq N_tE\left[-\log\left(\min_{u\in \mathcal{T}^{(m)}_{n}
}\epsilon_u\right)\mid |\mathcal{T}^{(m)}_{n}|\geq
d_{min}^{N_t-1}U-U^{\frac{1}{2}}\right]\left(1-U^{-\frac{1}{2}}\right),\quad U\geq
U_o.\nn
\end{equation}
 By applying Lemma~\ref{Lem:LogEps}
\begin{equation}
R^{(a)} \geq  \frac{N_t}{N_t-1}\left[\frac{\log M + \log
d_{min}^{N_t-1} + \log U + \log(1-U^{-\frac{1}{2}})}{P_{\alpha}} + \log
N_t\right]\left(1-U^{-\frac{1}{2}}\right),\quad U\geq U_o
\end{equation}
where $P_\alpha$ is modified from \eqref{Eq:Pa} as
\begin{equation}
P_{\alpha}=    1-\left[1-N_t2^{-(N_t-1)}\right]^{M(d_{min}^{N_t-1}U-U^{\frac{1}{2}})}
\end{equation}
  It follows from the
last inequality that
\begin{equation}
\lim_{U\rightarrow\infty}\frac{R^{(a)}}{\frac{N_t}{N_t-1}\log U}
\geq 1.\label{Eq:App:d}
\end{equation}
Combining \eqref{Eq:App:d} and \eqref{Eq:ThputUB:HiSNR} gives the
desired throughput scaling law for the interference-limited regime.

\section{Proof of
Proposition~\ref{Prop:Thput:LoSNR}}\label{App:Thput:LoSNR} From
\eqref{Eq:Thput:LoSNR} and since $0\leq \epsilon_u\leq 1$
\begin{equation}
R^{(\beta)} \leq  \E\left[\max_{1\leq m \leq M}\sum_{n=1}^{N_t}\log
\left(1+\gamma\max_{u\in\mathcal{I}^{(n)}_m}\rho_u\right)\right].
\end{equation}
In \eqref{Eq:App:k} in Appendix~\ref{App:ThputUB:NorSNR}, the above
upper-bound is also used for bounding the PU2RC throughput in the
normal SNR regime. Therefore, the upper-bound for the throughput scaling
factor as obtained in Appendix~\ref{App:ThputUB:NorSNR} is also
applicable for the present case, hence
\begin{equation}\label{Eq:App:i}
    \lim_{U\rightarrow\infty}\frac{R^{(\beta)}}{N_t\log\log U}\leq 1.
\end{equation}

Next, the above upper bond is shown to be achievable as follows. By
replacing the index set $\mathcal{I}^{(m)}_n$ in
\eqref{Eq:Thput:LoSNR} with its subset $\mathcal{T}^{(m)}_n$ defined
in \eqref{Eq:IndexSet:T}
\begin{eqnarray}
R^{(\beta)} &\geq& \E\left\{\sum_{n=1}^{N_t}\log
\left[1+\gamma\max_{u\in\mathcal{T}^{(n)}_m}\rho_u(1-\epsilon_u)\right]\right\}\nn\\
&\overset{(a)}{\geq}& \E\left\{\sum_{n=1}^{N_t}\log
\left[1+\gamma(1-d_{\min})\max_{u\in\mathcal{T}^{(n)}_m}\rho_u\right]\right\}\nn\\
&\overset{(b)}{\geq}& \E\left\{\sum_{n=1}^{N_t}\log
\left[1+\gamma(1-d_{\min})\max_{u\in\mathcal{T}^{(n)}_m}\rho_u\right]\mid |\mathcal{T}^{(n)}_m|\geq d_{\min}^{N_t-1}U - 1\right\}\left(1-\frac{1}{U}\right), U\geq U_o\nn\\
&\overset{(c)}{\geq}& N_t\E\left\{\log
\left[1+\gamma(1-d_{\min})\log(d_{\min}^{N_t-1}U-1)+\gamma(1-d_{\min})
O(\log\log
U)\right]\right\}\left(1-\frac{1}{U}\right)\times\nn\\
&&\left[1-O\left(\frac{1}{\log U}\right)\right]^{N_t}, U\geq U_o.\nn
\end{eqnarray}
The inequality (a) follows from the definition of
$\mathcal{T}^{(m)}_n$  in \eqref{Eq:IndexSet:T}. The inequality (b)
follows from \eqref{Eq:App:LargeNumb}. The inequality (c) is
obtained by using \eqref{Eq:ExtValTheo}. It follows from (c) that
\begin{equation}\label{Eq:App:j}
        \lim_{U\rightarrow\infty}\frac{R^{(\beta)}}{N_t\log\log U}\geq 1.
\end{equation}
Combining \eqref{Eq:App:i} and \eqref{Eq:App:j} gives the desired
throughput scaling law.

%============= References ====================================================
\linespread{1.6}
\bibliographystyle{ieeetr}

\newpage
%================ Tables ===========
%\begin{threeparttable}[h]
%  \centering
%    \caption{Comparison of PU2RC, OSDMA-S and OSDMA}\label{Tab:Design}
%  \begin{tabular}{r|ccc}
%    & PU2RC & OSDMA-S & OSDMA \\
%    \hline
%  Feeback/User (bits)~\tnote{a}& $2B + \logN$ & $B+I\logN_t$  & $B+\logN_t$  \\
%  Sum Capacity (bits/s/Hz)~\tnote{b} & largest (7.5) & moderate (6.4)  & smallest
%  (6.2)\\
%  Base Station Computation & largest & moderate & smallest \\
%  beam and user selection & centralized & distributed & N/A\\
%  User Conflict~\tnote{c} & No & Yes & Yes
%  \end{tabular}
%  \begin{tablenotes}\setlength{\baselineskip}{12pt}
%\item[a] {\footnotesize Assume $B$ bits are required for quantizing a channel gain and the quantization
%error of the channel shape .}
%\item[b] {\footnotesize Sum capacity is computed for $U=20$, $N_t=4$ and
%$\SNR=10dB$. Following \cite{ChoiForenza:OppSDMABeamSel:06}, the sum
%capacity is reduced by the feedback overhead factor $\lambda = 5\%$
%for each round of CSI feedback.}
%\item[c] {\footnotesize Refer to possibility that different users select a same beamforming vector.}
%\end{tablenotes}
%\label{Tab:AlgoCmp}
%\end{threeparttable}
%
%\vspace{40pt}
%================= Figures ============================

%------------------ Comparison with Non Asymptotic Scaling -------------------

%----------------- Effect of Increasing Channel Shape Feedback ---------------

%-------------- Compare ZF-SDMA --------------------------

%------------------------- Number of Users ---------------------------

%-------------- Compare ZF-SDMA variable SNR --------------------------

%------------- Compare DPC -----------------------------

%----------------- Effect of SINR Quantization ---------------

\end{document}